\documentclass[aps,pra,amsmath,amssymb,onecolumn,10pt,longbibliography,nofootinbib]{revtex4-1}
\usepackage[english]{babel}
\usepackage{xcolor,braket,enumitem,pifont,graphicx,accents,mathtools,bm}

\DeclareSymbolFont{symbolsC}{U}{txsyc}{m}{n}
\DeclareMathSymbol{\strictif}{\mathrel}{symbolsC}{74}
\DeclareMathSymbol{\boxright}{\mathrel}{symbolsC}{128}

\DeclareMathOperator{\erf}{erf}

\newcommand{\myvec}[1]{\accentset{\rightharpoonup}{#1}}

\newcommand{\cep}{\phi_\mathrm{CEP}}
\newcommand{\cepzero}{\phi_{\mathrm{CEP},0}}
\newcommand{\scep}{\sigma_\phi}
\newcommand{\tdet}{\tau_\mathrm{det}}
\newcommand{\pdet}{\myvec{p}_\mathrm{det}}
\newcommand{\deldet}{\myvec{\Delta}_\mathrm{det}}
\newcommand{\mean}[1]{\bar{#1}}
\newcommand{\tmin}{\tau_\mathrm{min}}
\newcommand{\tmax}{\tau_\mathrm{max}}
\newcommand{\drate}{\mean{d}}
\newcommand{\nh}{n_\mathrm{H}}
\newcommand{\nhtwo}{n_\mathrm{H2}}
\newcommand{\zdet}{\zeta}
\newcommand{\auxpars}{\{V_k\}}

\newcommand{\Edet}{E_\mathrm{det}}
\newcommand{\thmax}{\theta_\mathrm{max}}
\def \beq {\begin{equation}}
\def \eeq {\end{equation}}
\def \ba {\begin{eqnarray}}
\def \ea {\end{eqnarray}}

\newcommand{\simion}{\texttt{SIMION} }

\begin{document}

\title{Benchmarking strong-field ionisation with atomic hydrogen}

\author{D. Kielpinski$^{1,2}$, R. T. Sang$^{1,2}$, and I. V. Litvinyuk$^2$}

\affiliation{$^1$ARC Centre of Excellence for Coherent X-Ray Science and $^2$Australian Attosecond Science Facility, Centre for Quantum Dynamics, Griffith University, Nathan QLD 4111, Australia}

\begin{abstract}
As the simplest atomic system, the hydrogen atom plays a key benchmarking role in laser and quantum physics. Atomic hydrogen is a widely used atomic test system for theoretical calculations of strong-field ionization, since approximate theories can be directly compared to numerical solutions of the time-dependent Schr\"odinger equation. However, relatively little experimental data is available for comparison to these calculations, since atomic hydrogen sources are difficult to construct and use. We review the existing experimental results on strong-field ionization of atomic hydrogen in multi-cycle and few-cycle laser pulses. Quantitative agreement has been achieved between experiment and theoretical predictions at the 10\% uncertainty level, and has been used to develop an intensity calibration method with 1\% uncertainty. Such quantitative agreement can be used to certify experimental techniques as being free from systematic errors, guaranteeing the accuracy of data obtained on species other than H. We review the experimental and theoretical techniques that enable these results.
\end{abstract}

\date{\today}
\maketitle

\section{Introduction}

Strong-field ionisation is the starting point for experiments that reveal and control the dynamics of atoms, molecules, and solids \cite{Sheehy-DiMauro-high-field-rev, Corkum-Krausz-as-rev}. The interpretation of such experiments relies on an accurate understanding of the complex, highly nonlinear ionisation dynamics. The experimental data are often quite indirectly related to the physical parameters of interest and can easily show significant discrepancies from theoretical predictions. Retrieving accurate, quantitative information from strong-field experiments therefore requires a careful approach to the assessment and modeling of data. \\

Atomic hydrogen (H) offers unparalleled opportunities for benchmarking experimental and theoretical techniques in strong-field ionisation. As the simplest bound electronic system, H can be numerically simulated to extremely high precision using the exact time-dependent Schr\"odinger equation. Careful experiments on H can yield data that quantitatively agrees with the theoretical predictions to within experimental uncertainty. Both the data and the predictions are validated by such agreement. The validated predictions can then be used to calibrate any changes to the experimental parameters, while the validated data can certify the accuracy of the measurement technique for extension to future experiments. \\

\begin{figure}[thbp]
\begin{center}
\includegraphics[width=\columnwidth/2]{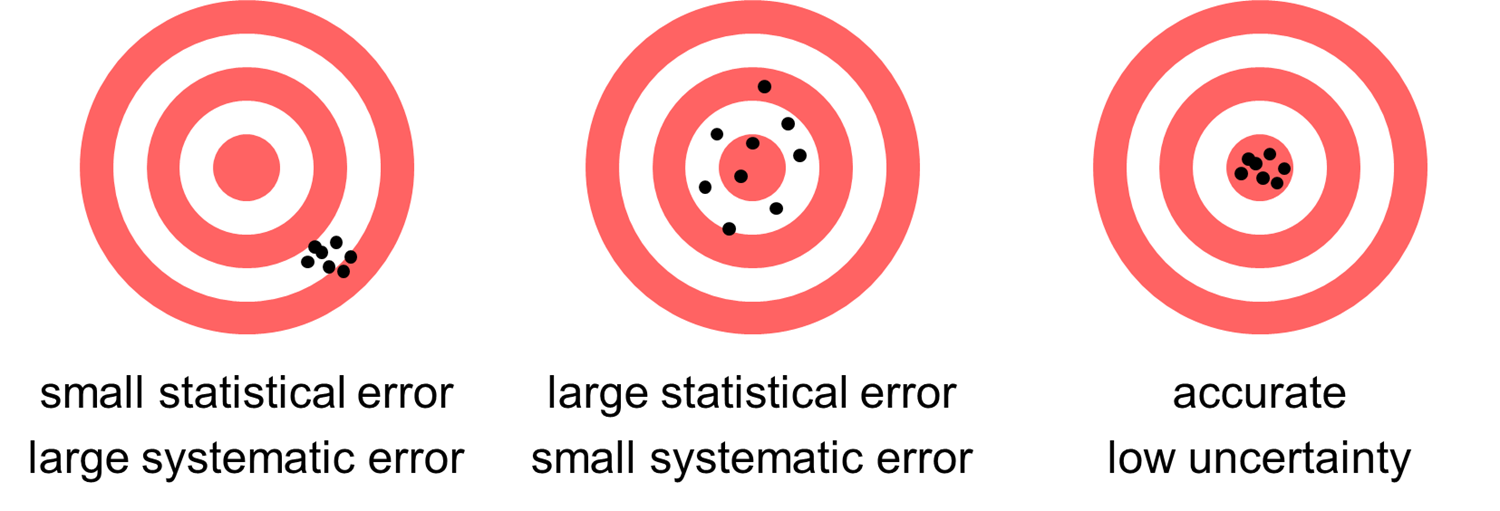}
\caption{Distinction between accuracy and precision. The left image shows small random error, i.e., high precision, but large systematic error. The middle image shows low-precision measurements that nevertheless have a small systematic error. The right image shows accurate measurements, which have both low systematic and low random error, and therefore have low uncertainty.}
\label{fig-Accuracy}
\end{center}
\end{figure}

The key benefit offered by experiments on H is \emph{accuracy}, as opposed to precision.\footnote{In this review, we use the terminology approved by the Bureau Internationale des Poids et Mesures (BIPM), the intergovernmental body responsible for metrology and precision measurement \cite{BIPM-international-vocabulary-metrology, BIPM-guide-uncertainty-measurement}.} Figure \ref{fig-Accuracy} illustrates the distinction between the concepts of accuracy and precision. Accuracy refers to the difference between the measured value and the true value of the quantity being measured. Precision refers simply to the difference between the values of repeated measurements. Even if all measurements agree with each other, they may still all differ from the true value. That difference is the systematic error of the measurement, as opposed to the random error associated with lack of precision. The total uncertainty of the measurement is derived from estimates of both systematic and random errors. For H, one can make essentially perfect theoretical predictions of the true values of measurement outcomes. If the measurement and the prediction agree to within the uncertainty of the measurement, we know that the systematic error is insignificant and the measurement is accurate as well as precise. \\

In this review, we discuss the experimental achievements in strong-field ionisation of atomic hydrogen and their relevance to benchmarking of strong-field physics. We summarise the experimental results that have been obtained so far with many-cycle, few-cycle, and carrier-envelope-phase stabilised laser pulses. Soon after the discovery of strong-field ionisation, experiments with many-cycle pulses played a key role in distinguishing resonantly enhanced, perturbative ionisation processes from true nonperturbative strong-field ionisation. In the few-cycle regime, highly accurate comparison of theory and experiment has been achieved with uncertainty at the 10\% level. This comparison has been used to measure few-cycle laser peak intensities with an uncertainty of 1\%. We detail the experimental and theoretical techniques that enable successful theory-experiment comparisons, as well as the statistical techniques needed for accurate estimation of the experimental uncertainty. While some elements of these techniques are well known to the community, this review brings together all the considerations relevant to state-of-the-art accuracy. \\

\section{Experiments with many-cycle pulses}

The first experiments on atomic hydrogen in intense laser pulses were performed by Rottke et al. \cite{Rottke-Welge-hydrogen-MPI} in the group of Welge at Bielefeld. These experiments used 500 fs pulses from an amplified dye laser system, tunable over 596 - 630 nm, to induce multiphoton ionisation. The small pulse bandwidth enabled Rottke et al. to observe electron spectral features with widths below 100 meV. Individual resonantly enhanced multiphoton processes could be identified through the excess kinetic energy imparted by the absorption of the last few photons by the excited atomic state. This interpretation was confirmed by measurements of the photoelectron angular distributions at the resonances. Later, Rottke et al. also observed higher-energy electron peaks (up to 12 eV), which they ascribed to multiphoton ionisation that was resonantly enhanced through the $4f$ excited state \cite{Rottke-Welge-hydrogen-low-order-ATI}. \\

Despite the quite high intensity achieved in these experiments, up to $1.2 \times 10^{14} \:\mbox{W}/\mbox{cm}^2$, the data obtained by Rottke et al. were readily interpreted within the Floquet theory of multiphoton processes \cite{Rottke-Shakeshaft-hydrogen-low-order-ATI}. Since the ponderomotive energy $U_p$ scales as $\lambda^2$, the ponderomotive energy in these experiments was about half of that for an equivalent intensity of 800 nm light, so that $U_p \lesssim 4$ eV. The Keldysh parameter for ionisation from the ground state was therefore $\gamma \equiv \sqrt{I_p/(2 U_p)} \gtrsim 1.2$, where $I_p = 13.6$ eV is the ionisation energy of H. Naively, one would expect tunnelling physics to be relevant, if not dominant, in this regime. However, for the low electron energies examined in this experiment ($< 12$ eV), the electron yield was strongly resonantly enhanced by the highly excited $4f$ state and other similarly high-lying states. The binding energy of the $4f$ state is only $E_{4f} = 0.85$ eV, so the final ionisation step was far in the barrier suppression regime (in this case, $I \gtrsim 10^{10} \:\mbox{W}/\mbox{cm}^2$). Hence, tunneling physics played a minor role in the interpretation of these results. \\

\begin{figure}[thbp]
\begin{center}
\includegraphics[width=\columnwidth/2]{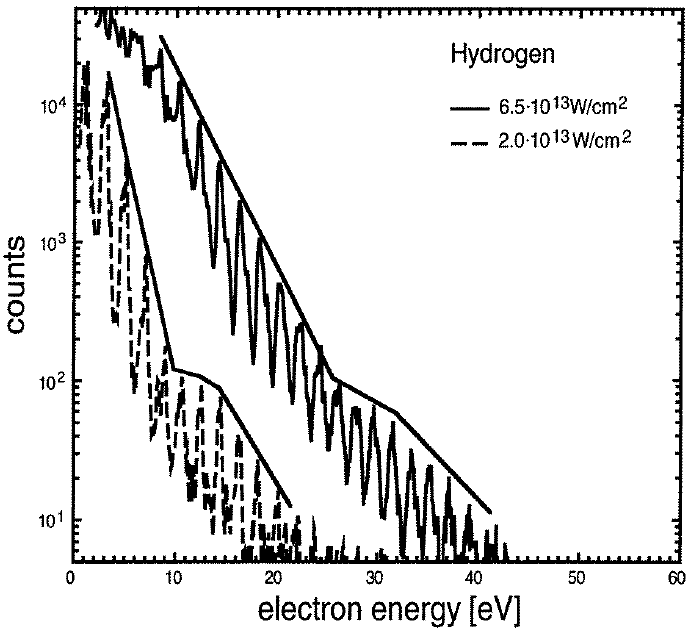}
\caption{Observation of strong-field ionisation in atomic deuterium \cite{Paulus-Walther-hydrogen-ATI-expt}. The slanted lines are guides to the eye. A change in slope is clearly observed, indicating the presence of a plateau.}
\label{H_long_pulse}
\end{center}
\end{figure}

Experimental investigations of true strong-field physics in this area were initiated by Paulus et al. \cite{Paulus-Walther-hydrogen-ATI-expt} in the group of Walther at the Max Planck Institute for Quantum Optics. These experiments examined electron energies up to 40 eV and were the first to observe the ATI plateau in a hydrogen-like system (Figure \ref{H_long_pulse}). Paulus et al. actually used deuterium in order to maximise the efficiency of their molecular dissociation source \cite{Paulus-Walther-hydrogen-ATI-expt}. However, the deuterium atom is still a one-electron, one-nucleus system and the nuclear mass has little effect on strong-field ionisation, so deuterium and hydrogen are expected to exhibit virtually identical photoelectron spectra. Again, visible pulses from an amplified dye laser system were used, with wavelength of 630 nm, but with a much shorter pulse duration of 40 fs. The intensities were similar to those of Rottke et al., up to $0.7 \times 10^{14} \:\mbox{W}/\mbox{cm}^2$. \\

From these experiments, Paulus et al. were able to demonstrate that electrons with energies above $\sim 12$ eV originated in strong-field ionisation from the ground state. Plateau features were clearly seen in simulated spectra from numerical solutions of the time-dependent Schr\"odinger equation, and the energy range of the plateau corresponded to the experimentally observed features. However, as often found in strong-field ionisation experiments, the theoretically predicted strong plateau was significantly obscured by the effect of the inhomogeneous laser intensity, the so-called ``focal volume integration''. This crucial effect, which is central to any comparison of experiment and theory in strong-field ionisation, is discussed in detail in Section \ref{sec-compare}. \\

The comparison of Paulus et al's data with the work of Rottke et al. provided further evidence for nonperturbative strong-field dynamics. Resonantly enhanced multiphoton processes were still clearly observed at electron energies of a few eV. However, the plateau peaks were only observed at the higher intensities, indicating the appearance of nonperturbative dynamics as the Keldysh parameter was decreased from $\gamma \approx 2$ to its lowest value of $\gamma = 1.6$. The low-energy electron peaks were also shifted by 0.8 eV relative to the plateau peaks, indicating that they originate from a separate ionisation process. \\

\section{Experiments with few-cycle pulses}
\label{sec-aasfexpts}

Experiments on the ionisation of H with few-cycle pulses commenced in 2009 at the Australian Attosecond Science Facility (AASF) at Griffith University. These experiments have attempted to realise the potential of H to benchmark strong-field ionisation. Few-cycle ionisation and carrier-envelope phase effects have been the main focus, since the physics of these phenomena is highly relevant to attosecond science. The initial goal was to achieve agreement between theory and experiment to within a small, accurately estimated margin of error. Achieving this goal has led to validation of experimental techniques and to novel calibration methods for strong-field experiments. \\

\subsection{Experimental apparatus}
\label{sec-apparatus}

\begin{figure}[thbp]
\begin{center}
\includegraphics[width=\columnwidth/2]{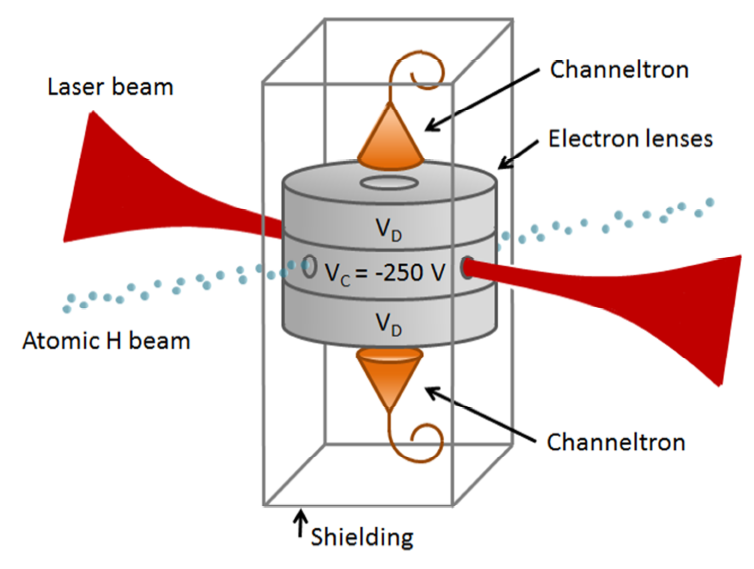}
\caption{Experimental schematic for few-cycle photoionisation experiments on atomic H.}
\label{fig-exptschem}
\end{center}
\end{figure}

The apparatus used for the AASF experiments is schematically depicted in Figure \ref{fig-exptschem}. A few-cycle laser beam intersects an atomic beam of hydrogen. The photoelectrons enter a repeller system, which only passes electrons above a cutoff energy. The high-energy electrons are counted by a channeltron detector. We give a brief review of the experimental design below: further details can be found in \cite{Pullen-Kielpinski-few-cycle-H-ionisation, Pullen-Kielpinski-H-ionisation-THESIS, Wallace-Kielpinski-H-ionisation-CEP-effects}. \\

\subsubsection{Atomic hydrogen target}
\label{sec-htarget}

The preparation of atomic hydrogen is, in principle, straightforward. $\mbox{H}_2$ gas is dissociated by electron impact in an electrical discharge or simply by heating to a temperature above the dissociation energy. However, the dissociation process is not particularly efficient, and atomic H readily recombines at most surfaces. For precision measurements, one requires a well-characterised and relatively pure source of H. \\

The AASF experiments use a radiofrequency (RF) discharge dissociator based on the design of Slevin \cite{Slevin-Stirling-rf-H-dissociator}. The dissociator is essentially identical to those used in the group of Kleppner \cite{Schwonek-Kleppner-H-source-THESIS}. $\mbox{H}_2$ gas at a backing pressure of $\sim 2$ mbar flows through a Pyrex tube. A discharge is struck within the tube, driven by the high-voltage output of a helical RF resonator operating at 75 MHz. Approximately 10 W of RF power is dissipated in the tube, and water cooling is used to keep the tube temperature below 20 C. The gas that emerges from the discharge tube consists of up to 90\% atomic H by number. \\

The requirements for efficient operation of the H source pose substantial technical difficulties for strong-field ionisation experiments. The outlet of the discharge tube has a 4 mm inner diameter to minimise recombination of the atomic H, much wider than the typical gas-jet nozzle. The gas load of the source is therefore very high, on the order of $5 \times 10^{-3} \:\mbox{mbar L}/\mbox{s}$. Recombination of this high gas load at the chamber walls creates a large background pressure of $\mbox{H}_2$. At the same time, the backing pressure is approximately $100\times$ lower than in a typical gas-jet experiment, so the atomic density is low. \\

To isolate the H atoms from the $\mbox{H}_2$ background, the gas from the source passes from the source chamber, through a few mm aperture into a differential pumping chamber, then through a 0.5 mm aperture into the experiment chamber. The entire system is $\sim 2$ meters long, much shorter than the mean free path under the vacuum operating conditions, and the apertures are separated by $\sim 1$ meter. Hence the atoms that emerge into the experiment chamber form a well-collimated atomic beam, while very little of the $\mbox{H}_2$ gas from the previous chambers escapes into the experiment chamber. However, the number density of the atomic beam is extremely low. Even at the operating pressure of $3 \times 10^{-10}$ mbar, the background gas in the experiment chamber (mostly $\mbox{H}_2\mbox{O}$ and $\mbox{H}_2$) makes a significant contribution to the photoionisation signal. \\

Precision measurements of photoionisation from H require the subtraction of photoionisation signals that originate from the background gas and from undissociated $\mbox{H}_2$ in the atomic beam. To this end, we perform three sets of measurements: (1) with the discharge turned on, obtaining a signal $C_\mathrm{ON}$, (2) with the discharge turned off, obtaining $C_\mathrm{OFF}$, (3) with the atomic beam blocked, obtaining $C_\mathrm{BACK}$. A pneumatically actuated gate valve between the source chamber and the differential pumping chamber serves to block the atomic beam. The conditions for the three measurements are otherwise made as identical as possible. The signal due to H can then be computed as \cite{Pullen-Kielpinski-few-cycle-H-ionisation, Pullen-Kielpinski-H-ionisation-THESIS}
\beq
S_H \propto (C_\mathrm{ON} - C_\mathrm{BACK}) - (1 - \mu) (C_\mathrm{OFF} - C_\mathrm{BACK}) \label{subtrax}
\eeq
where $\mu = \nh/(\nh + 2\nhtwo)$ is the mass dissociation fraction of the source. Equation (\ref{subtrax}) shows that we must know the value of $\mu$ in order to make an accurate measurement of $S_H$. \\

We determine the value of $\mu$ with 10\% uncertainty through emission spectroscopy of the RF discharge \cite{Lavrov-Ropcke-emission-spectrum-dissoc-fraction-theory, Lavrov-Ropcke-emission-spectrum-dissoc-fraction-expt}. In this technique, the light emitted from the discharge is collected and the strengths of the Balmer-$\alpha$ and $\beta$ lines of atomic hydrogen and several lines in the Fulcher-$\alpha$ band of $\mbox{H}_{2}$ are measured with a carefully calibrated monochromator. The ratios of line strengths are compared to a rate-equation model of the radiative and collisional processes in the plasma, allowing the simultaneous determination of $\mu$, the electron temperature of the plasma, and the gas temperature of the plasma. The model has been verified over a wide variety of discharge conditions \cite{Lavrov-Ropcke-emission-spectrum-dissoc-fraction-expt}. Details of our implementation can be found in \cite{Pullen-Kielpinski-H-ionisation-THESIS}. \\

\subsubsection{Laser system}
\label{sec-laser}

The laser system used in the AASF hydrogen experiments is a Femtopower Compact Pro CE-Phase (Femtolasers GmbH). This system, which is now quite standard for few-cycle experiments, produces pulses of $\sim 6$ fs duration at 800 nm wavelength. The pulse energy at the target is $\sim 300 \:\mu$J, and it is relatively straightforward to focus the pulses to a $1/e^2$ beam radius of $\sim 40 \:\mu$m using an off-axis paraboloidal mirror, thus achieving intensities up to $10^{15} \:\mbox{W}/\mbox{cm}^2$. \\

The laser system consists of a titanium:sapphire (Ti:S) few-cycle oscillator, chirped-pulse Ti:S amplification, and a hollow-fibre compressor. The oscillator produces $\sim 6$ fs pulses at 80 MHz. Dispersion-compensating mirrors are used to achieve near-zero cavity dispersion in the oscillator. Spectral broadening and difference-frequency generation of the oscillator pulses is used to generate an $f-0f$ carrier-envelope beat signal, which is actively stabilised to $1/4$ of the laser repetition rate. After stretching, the oscillator pulses enter a multipass Ti:S amplifier with a repetition rate of 1 kHz and are then recompressed in a prism compressor to $\sim 26$ fs duration. The amplified pulses undergo spectral broadening in a glass capillary waveguide (''hollow fibre'') of $\sim 1$ m length and $250 \:\mu$m inner diameter. The spectrally broadened pulses are then recompressed with a dispersion-compensating mirror set to achieve a final pulse duration of $5.5 - 6.3$ fs, depending on the day-to-day laser stability and the expertise of the operator. A commercial $f-2f$ interferometer is used after the hollow fibre to measure and stabilise the CEP of the amplified laser pulses. The laser beam is focused into the experiment chamber with an off-axis paraboloidal mirror of 750 mm focal length. \\

At the experimental station, the laser intensity is controlled by transmission through a set of uncoated pellicle membranes on flip mounts. The laser power, and thus intensity, is reduced by Fresnel reflections from each pellicle. The standard technique for controlling the laser power, namely, rotating the laser polarisation with a half-wave plate and then passing the laser through a polariser, appears inadequate for precision measurements with few-cycle pulses. Measurements at the AASF have shown significant deviation from 3D-TDSE predictions when using this technique, whereas the pellicle technique is highly reliable. Optically coated pellicles are found to be suitable for CEP-averaged measurements, but introduce unpredictable changes to the CEP and are therefore unsuitable for comparing CEP-dependent measurements at different intensities. The use of coated glass beamsplitters is also not suitable, since different beamsplitters are found to introduce slightly different wavefront curvature to the beam, thus slightly changing the laser spot size. \\

The laser beam parameters are carefully measured during the experiment. The laser passes through a low-reflectivity beamsplitter, providing a sample beam for measurement of pulse duration. The beam sample passes through a thickness of fused silica to add a group delay dispersion that is nominally equal to that experienced by the main beam, and its pulse duration is measured by a few-cycle autocorrelator. Another sample beam is picked off after the paraboloidal mirror and sent to a camera-based beam profiler for monitoring of focused spot size and beam quality. The laser power is monitored by a thermal power meter and/or photodiode. \\

\subsubsection{Repeller detector}
\label{sec-repeller}

\begin{figure}[thbp]
\begin{center}
\includegraphics[width=\columnwidth/2]{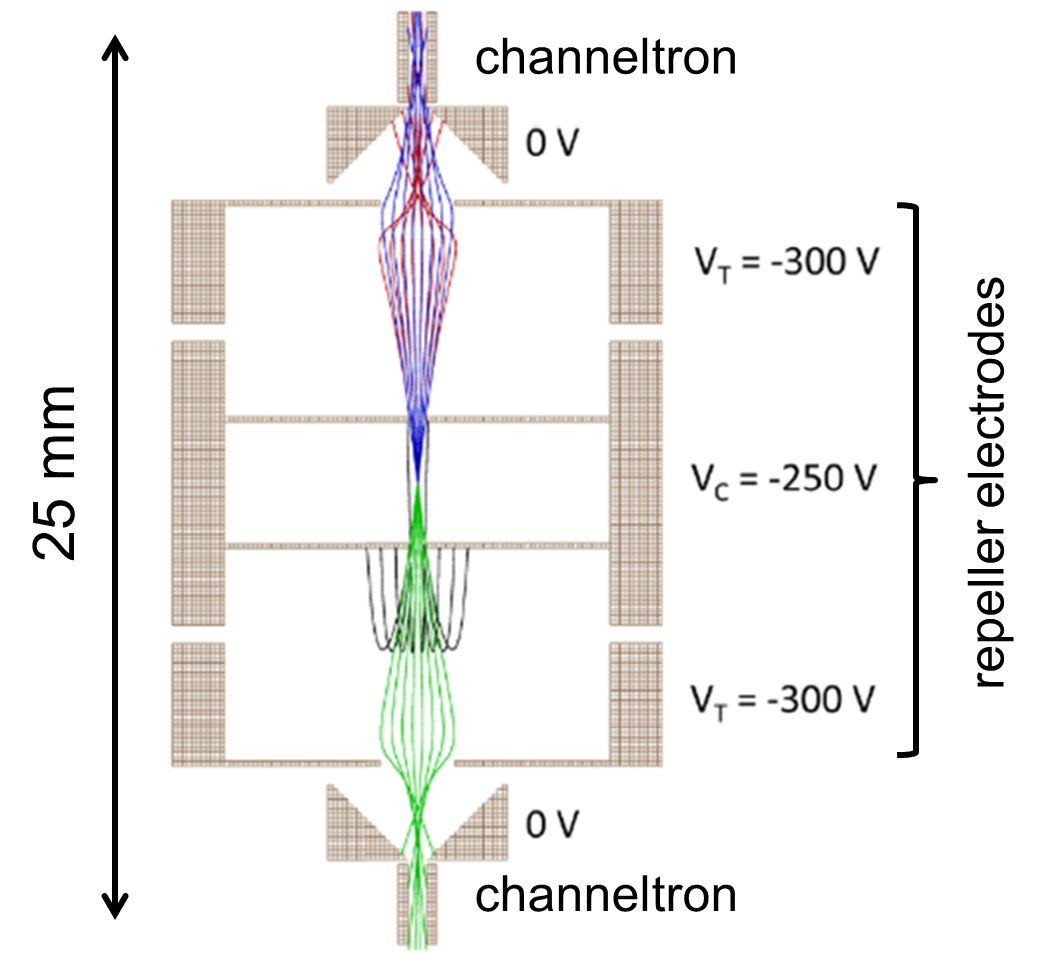}
\caption{Repeller detector. Electrons originate from photoionisation at the centre of the detector assembly. High-energy electrons overcome the potential barrier between $V_C$ and $V_T$ and are detected at the channeltrons. Electron trajectories (colored lines) are simulated with \simion. Trajectories with initial energy of 25 eV (black lines) do not overcome the barrier, but trajectories with energies of 40 (red), 50 (green), and 80 eV (blue) reach the channeltrons and are detected. }
\label{fig-detector}
\end{center}
\end{figure}

The detection system consists of a set of ``repeller'' electrodes that pass only the high-energy electrons, two channel electron multipliers (``channeltron'') for electron detection, and standard electronics for channeltron signal amplification and discrimination. Figure \ref{fig-detector} shows the geometry of the repeller electrodes and the channeltrons. We fix the voltage $V_C$ to $-250$ V and vary $V_T$ in the range $-250$ to $-350$ V, creating a potential barrier that repels the low-energy electrons. The channeltrons are grounded, and the 250 -- 350 V acceleration from $V_T$ to the channeltrons ensures that the quantum efficiency of the channeltron detection is nearly independent of electron energy. \\

The commercial charged-particle optics package \simion was used to simulate electron trajectories for various initial conditions and repeller electrode voltages. These simulations show that the only electrons that reach the channeltron are those with initial energy greater than some cutoff energy $E_c$. The cutoff energy depends on the ``deflection potential'' $V_D = V_C - V_T$ as $E_c \approx 0.8(V_D - 5)$, where $E_c$ is measured in eV and $V_D$ is measured in V. \\

For a given cutoff energy $E_c$, the number of electrons $Y(E_c)$ detected at one of the channeltrons is approximately proportional to the integral of the electron energy spectrum $S(E)$
\beq
Y(E_c) \propto \int_{E_c}^{\infty} \!dE\: S(E)
\eeq
The justification for this approximation and the correction for systematic errors in this detector model are discussed in Section \ref{sec-detector-model}.

\subsection{Photoelectron measurements}
\label{sec-elec-meas}

Highly accurate measurements of the few-cycle photoelectron yield $Y(E_c,I)$ as a function of both electron cutoff energy $E_c$ and laser peak intensity $I$ were performed at the AASF in 2010 \cite{Pullen-Kielpinski-few-cycle-H-ionisation, Pullen-Kielpinski-H-ionisation-THESIS}. For these measurements, the laser pulse duration was 6.3 fs and the laser CEP was made to vary rapidly, thus averaging over a uniform CEP distribution. Over 250 data points were obtained, spanning an order of magnitude in energy (5 - 60 eV) and intensity (1.2 - 5.4 $\times 10^{14} \:\mbox{W}/\mbox{cm}^2$). Statistical analysis of the data and estimation of uncertainty was performed using the techniques discussed below, in Section \ref{sec-accurate-meast}. Detailed modelling of the experiment enabled the derivation of realistic predictions of the data from the theoretical simulations, according to the principles of Section \ref{sec-compare}. In fitting the predictions to the experiment, only two fit parameters were used for the entire dataset. Given a theoretical prediction $Y_\mathrm{pred}(V_D,I)$, we performed a least-squares fit to the entire dataset $Y(V_D,I)$, using the fit function
\beq
Y_\mathrm{fit}(V_D,I_\mathrm{nom};A,\eta) = A \: Y_\mathrm{pred}(V_D, \eta I_\mathrm{nom}) \label{globalfit}
\eeq
with $A$ and $\eta$ as fit parameters and $I_\mathrm{nom}$ as the nominal peak intensity, measured from the laser power, pulse duration, and spot size. The fit parameter $\eta$ can be interpreted as an overall rescaling of the absolute laser intensity, while $A$ accounts for the absolute quantum efficiency of the detection system and the unknown absolute target density. In the fits, $V_D$ was employed rather than $E_c$ so as to account for any slight deviations from the relation $E_c \approx 0.8(V_T - 5)$. For ease in interpretation, the independent variable was then changed from $V_D$ to $E_c$ using that relation. \\

\begin{figure}[thbp]
\begin{center}
\includegraphics[width=\columnwidth]{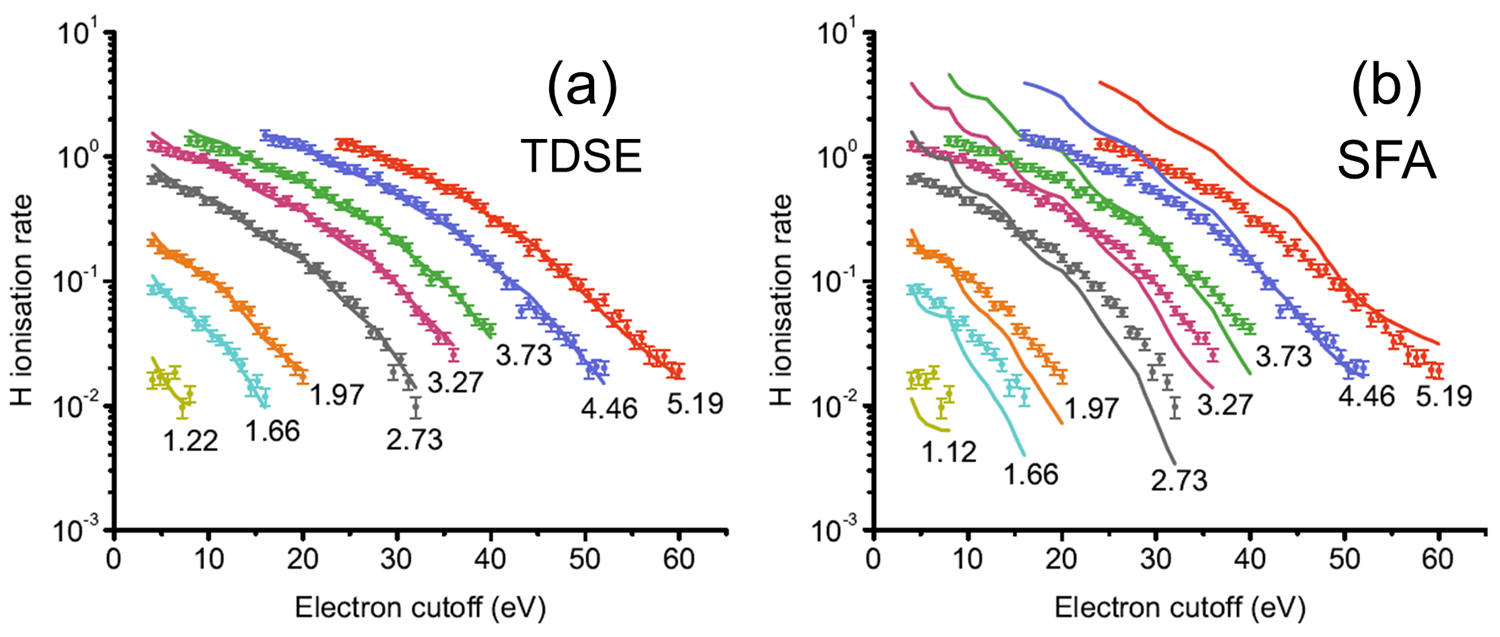}
\caption{Few-cycle photoelectron yield measurements in H and comparison to theory. Points with error bars: data. Lines: theoretical predictions. The numerical label on each data series gives the intensity in units of $10^{14} \:\mbox{W}/\mbox{cm}^2$. (a) Comparison to TDSE prediction. (b) Comparison to SFA prediction.}
\label{fig-EYields}
\end{center}
\end{figure}

Figure \ref{fig-EYields}(a) shows the remarkable agreement between experiment and a prediction based on exact 3D time-dependent Schr\"odinger equation (3D-TDSE) simulations. The uncertainty in each of the data points was approximately $10\%$, and again, only two fit parameters were used for all the points, more than 250 in total. Nevertheless, the data and predictions agreed everywhere within the estimated experimental uncertainty. Careful modelling of the experiment was crucial for obtaining this level of agreement. For instance, focal volume averaging effects (Section \ref{sec-fvi}) altered the predicted yield curve by $\sim 50\%$, far more than the experimental uncertainty. For the 3D-TDSE prediction, the intensity rescaling factor $\eta$ was determined to be $0.94 \pm 0.01$, well within the usual uncertainty of the method used to measure $I_\mathrm{nom}$ \cite{Pullen-Kielpinski-peak-intensity-calibration}. \\

In contrast, predictions based on the strong-field approximation (SFA) were sharply ruled out by the data, as shown in Figure \ref{fig-EYields}(b). Since the data was validated by the agreement with 3D-TDSE predictions, the disagreement with the SFA showed that the SFA did not provide adequate predictions in this case. Not only did the prediction fall far outside the experimental uncertainty, but the retrieved value of the intensity factor $\eta$ was also found to be $1.33 \pm 0.05$, far outside the uncertainty of $I_\mathrm{nom}$. As discussed in the Introduction, retrieving reliable physical parameters requires accurate agreement between prediction and experiment. Since the SFA prediction was not adequately accurate, it is not surprising that the SFA retrieved an unreliable value of $\eta$. \\

\subsection{Carrier-envelope phase effects}

Experiments to accurately measure CEP effects on the photoelectron yield were undertaken at the AASF in 2011 - 2012 \cite{Wallace-Kielpinski-H-ionisation-CEP-effects}. For these experiments, the laser pulse duration was slightly shorter (5.5 fs). The electron yield $Y_{T,B}(\cep,E_c,I)$ was measured for both top ($Y_T$) and bottom ($Y_B$) detectors as a function of CEP, $\cep$, for a range of electron cutoff energies $E_c$ and laser peak intensities $I$. At each fixed value of $E_c$ and $I$, the value of $\cep$ was stepped and the yield was measured for each $\cep$ value. The CEP was stepped by changing the insertion of a fused silica wedge pair in the laser beam. Stepping the CEP electronically, via the laser stabilisation system, was found to produce unreliable results. \\

Following standard practice \cite{Rathje-Sayler-ATI-CEP-measurement-rev}, the CEP effect at a given value of $E_c$ and $I$ was parametrised through the asymmetry, defined by
\ba
A(\cep) &=& \frac{Y_T(\cep) - Y_B(\cep)}{Y_T(\cep) + Y_B(\cep)} \\
&\approx& A_0 \sin (\cep - \phi_0) \label{cep-sin-fit}
\ea
The asymmetry was found to follow a sinusoidal function of period $2 \pi$ within the uncertainty of the individual data points, as expressed in the second equality. Fitting the measured $A(\cep)$ to this sinusoid yielded the asymmetry amplitude $A_0$ and phase offset $\phi_0$. A map of the amplitude $A_0(E_c,I)$ was constructed, covering a wide range of electron cutoff energies $E_c$ and laser peak intensities $I$. A map of the phase offset was also constructed \cite{Wallace-Kielpinski-H-ionisation-CEP-effects}, but its precision was not high enough to admit useful interpretation and we omit it here. \\

\begin{figure}[thbp]
\begin{center}
\includegraphics[width=\columnwidth]{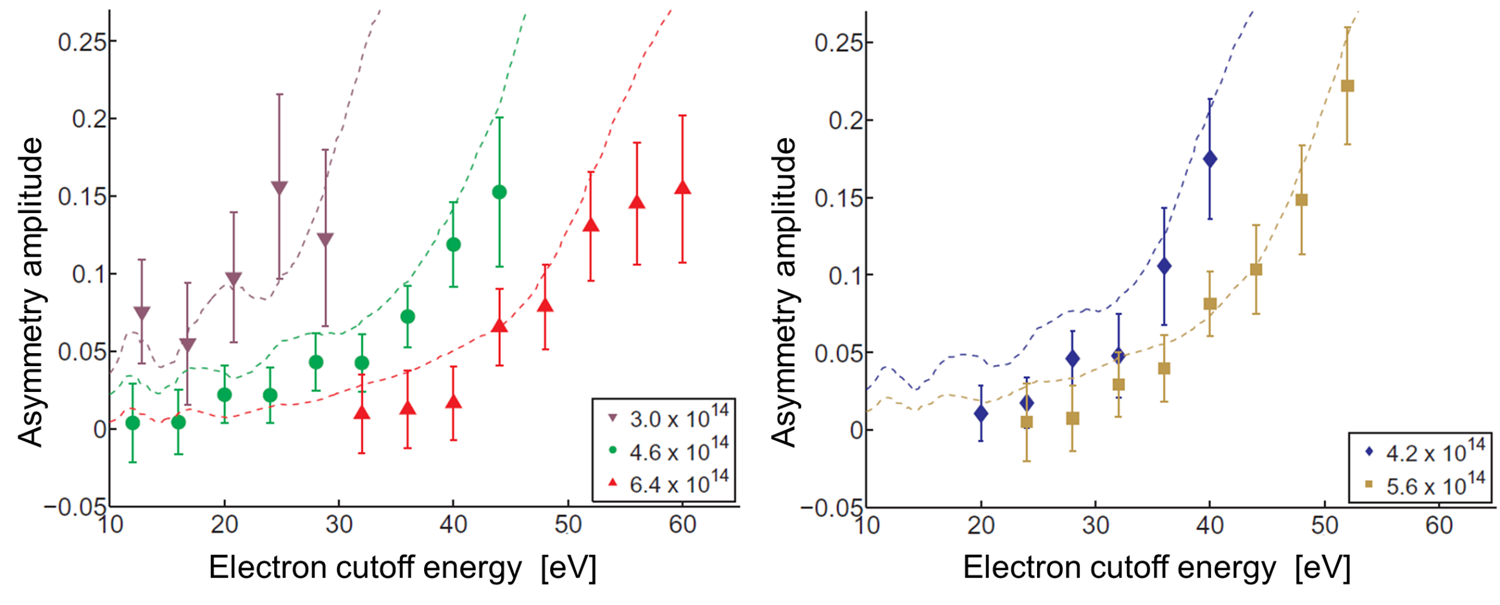}
\caption{CEP effects on few-cycle photoelectron yield in H. The asymmetry amplitude $A_0$ is shown as a function of electron energy $E_c$ over a range of intensities (labeled in legends, in units of $10^{14} \mbox{W}/\mbox{cm}^2$.}
\label{fig-PhaseEffects}
\end{center}
\end{figure}

Figure \ref{fig-PhaseEffects} shows the map of asymmetry amplitude $A_0(E_c,I)$ and compares it to the predictions derived from 3D-TDSE simulations. The data and predictions agreed \emph{with no free parameters} over the entire range of $E_c$ and $I$. Again, careful modelling of the experiment was essential to accurately predict the observations. In particular, it was found that the 10 meter air path between the laser system and the experiment chamber introduced large fluctuations of the CEP. Measurement of the CEP with an $f-2f$ interferometer at the experiment chamber showed a root-mean-square uncertainty of 850 mrad in the CEP, with a corresponding reduction in the asymmetry amplitude per Section \ref{sec-cep-avg}. This CEP instability also showed why the simple sinusoidal fit, Eq. (\ref{cep-sin-fit}), was adequate to model the data. \\

The asymmetry amplitude $A_0$ was seen to increase rapidly as the electron cutoff energy $E_c$ approached $2 U_p$, a trend that could be systematically observed for all intensities. This trend accords with theoretical expectations for the behaviour of high-energy direct electrons \cite{Paulus-Becker-CEP-depdt-ionisation-rev}. It was not possible to observe the plateau electrons, since the H beam density was too low to provide adequate signal. \\

\subsection{Intensity calibration standard}

The excellent agreement between experiment and theory for H enables highly accurate retrieval of experimental parameters. By obtaining experiment-theory agreement over a wide parameter range, one establishes that the theoretical predictions reliably reflect the experimental conditions. Alternatively, such agreement can be interpreted as showing that the experimental apparatus is accurately calibrated. Later on, if more data are obtained under imperfectly known experimental conditions, these data can be fitted to theory. So long as the fit between theory and data remains within the known uncertainty of the data, the fit coefficients will reliably estimate the experimental parameters. \\

The laser peak intensity represents a particularly important target for calibration and estimation. Intensity changes of a few percent can radically alter strong-field ionisation \cite{Hansch-VanWoerkom-ATI-fast-intensity-dependence, Paulus-Becker-channel-closing}. However, the intensity of strong-field laser pulses is difficult to measure directly. One can combine measurements of laser power (via, e.g., power meter), pulse duration (autocorrelator), and spot size (camera-based beam profiler) to obtain an intensity measurement. The uncertainty of this technique is still approximately 10\% even when care is taken with the measurements, mostly owing to uncertainty of the laser spot size \cite{Pullen-Kielpinski-peak-intensity-calibration}. Moreover, there may be systematic differences between the measured beam properties and the actual beam properties at the interaction region. \\

Several methods for \emph{in situ} intensity calibration have been developed, but while these methods offer high precision, they are all vulnerable to uncontrolled systematic errors. The total photoionisation yield may be measured at several estimated intensities and compared to theoretical predictions for the yield \cite{Larochelle-Chin-ADK-ionisation-yield-comparison}. The photoelectron and photoion momentum distribution may be compared to semiclassical theory \cite{Litvinyuk-Corkum-aligned-molecule-ionisation, Alnaser-Cocke-peak-intensity-calibration, Smeenk-Staudte-peak-intensity-calibration}. Finally, the photoelectron spectrum may be compared to quantitative rescattering theory \cite{Micheau-Lin-laser-parameter-photoelectron-retrieval, Chen-Lin-CEP-ATI-waveform-retrieval}. In each of these experiments, the data were compared to an approximate theory, but it was not possible to estimate the effect of the approximations on the intensity retrieval, since exact theories are not available for species other than H. \\

The AASF experiments on H provided intensity calibration with total uncertainty below 1\% in the range from 1 to 5 $\times 10^{14} \mbox{W}/\mbox{cm}^2$ \cite{Pullen-Kielpinski-peak-intensity-calibration}. There were no systematic errors arising from theoretical approximations, since the accuracy of the calibration was guaranteed by the agreement between data and 3D TDSE predictions. These experiments also showed that introducing even reasonable and common approximations to the theory can induce errors on the 10\% level, comparable to the uncertainty of intensity estimates derived from measuring laser beam properties. \\

To calibrate the intensity, the AASF team acquired data on the photoelectron yield $Y_\mathrm{cal}(V_D)$ as a function of deflector voltage $V_D$, keeping the laser intensity constant. The data was fitted to the function
\beq
Y_\mathrm{fit}(V_D;A,I_\mathrm{TDSE}) = A \: Y_\mathrm{TDSE}(V_D,I_\mathrm{TDSE}) \label{singlefit}
\eeq
where $Y_\mathrm{TDSE}(V_D,I)$ is the predicted yield based on TDSE simulations. The fit parameters were now $I_\mathrm{TDSE}$, the laser peak intensity as retrieved by the TDSE, and $A$, the detection efficiency factor. The fit returned the peak intensity for the dataset. \\

The intensity calibration procedure should be carefully distinguished from the test of experiment-theory agreement in Section \ref{sec-compare}. In the agreement test, the entire dataset $Y(V_D,I)$, ranging over all intensities, was fitted and found to agree with theory. The global intensity rescaling factor accounted for our uncertainty in measuring the intensity, thus enabling a good match between experiment and theory. For intensity calibration, we \emph{relied} on the global agreement between experiment and theory that was already established in Section \ref{sec-compare}. Given this global agreement, we could be confident that new data $Y_\mathrm{cal}(V_D)$ would also agree with the TDSE predictions under an appropriate fitting procedure. Agreement of the new data with the TDSE predictions would occur precisely when the fitted intensity correctly estimated the true (unknown) intensity. \\

\begin{figure}[thbp]
\begin{center}
\includegraphics[width=\columnwidth/2]{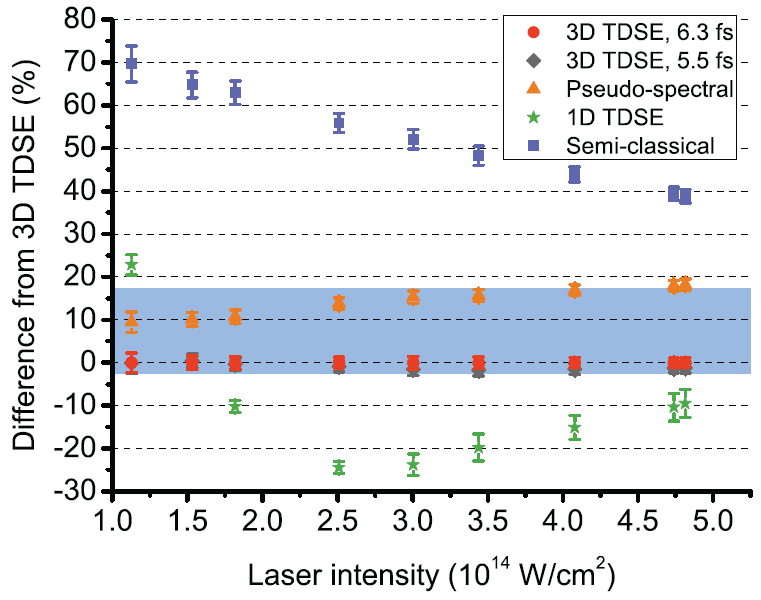}
\caption{Comparison of theoretical predictions for use in intensity calibration. For each theory, the fractional difference $\epsilon_X$ from the 6.3 fs 3D TDSE prediction is shown as a function of the intensity retrieved by the 6.3 fs 3D TDSE. See Eq. (\ref{intenserr}) for definition of $\epsilon_X$. Gray bar: uncertainty in intensity for independent measurements of laser beam parameters. }
\label{fig-IntensCal}
\end{center}
\end{figure}

Unlike the 3D TDSE, approximate theories were found to be incapable of accurately calibrating the intensity. The fractional systematic error in theory X (e.g., 1D TDSE theory) is given by
\beq
\epsilon_X = \frac{I_\mathrm{est,X} - I_\mathrm{TDSE}}{I_\mathrm{TDSE}} \label{intenserr}
\eeq
where $I_\mathrm{est,X}$ is the intensity estimated using the prediction of theory X. Naturally, $\epsilon_X$ varied depending on the actual intensity $I_\mathrm{TDSE}$. Figure \ref{fig-IntensCal} shows the systematic error $\epsilon_X(I_\mathrm{TDSE})$ for a selection of approximate theories, measured relative to the intensity estimated by the 3D TDSE prediction at 6.3 fs pulse duration. Details of the theoretical methods are given in \cite{Pullen-Kielpinski-peak-intensity-calibration}. The agreement of the two TDSE simulations, for 5.5 and 6.3 fs pulse durations, showed that the intensity calibration technique was relatively insensitive to the pulse duration. However, for all theories except for the 3D TDSE, $\epsilon_X$ fell outside the 10\% uncertainty range obtained by independent measurements of the laser beam parameters. Only the \emph{in situ} calibration that uses the TDSE could be regarded as an improvement over an intensity estimate based on measurement of the laser beam parameters.  \\

\section{Principles for analysing strong-field ionisation data}
\label{sec-accurate-meast}

\subsection{Estimating particle arrivals from detector signals}

Charged-particle detectors work by amplifying the electrical current received at the detector input due to the arrival of the particles. Multichannel plates are the most commonly used detectors in strong-field ionisation experiments, though channel electron multipliers are also used. Since most signal-processing electronics uses voltage rather than current, one normally converts the detector output current to a voltage using a resistor or a transimpedance amplifier. \\

Present-day detectors amplify so strongly that the detection of a single particle results in a readily detectable voltage pulse. However, a certain fraction of particle arrivals simply fail to register at the detector at all. The fraction of arrivals that result in a voltage pulse is called the ``quantum efficiency'' of the detector. It is straightforward to ensure that the quantum efficiency is independent of particle momentum. On the other hand, when no particles are arriving, the voltage fluctuations are much smaller than the typical pulse resulting from a particle detection. \\

Detectors are commonly operated either in counting mode or in proportional mode. In counting mode, the successful detection of a particle gives rise to short voltage pulses on all the readouts $V_j$. As long as the time between detection events is much longer than the voltage pulse duration $\tdet$, each detection event can be readily distinguished and ``counted''. In proportional mode, one integrates the entire voltage waveform $V_j(t)$ over an appropriate time interval. The integral is nominally proportional to the number of detected particles. \\

Counting-mode detection is highly immune to electronic noise. For instance, suppose we wish to count the number of electrons released by a laser pulse in order to make a photoelectron yield measurement. We set up a simple detector with a single voltage readout $V(t)$, which rises to a peak voltage $V_0$ when a particle is successfully detected. To count particles, we count the number of times that $V(t)$ rises above a fixed discriminator voltage $V_\mathrm{disc} < V_0$. Whenever this happens, we register a ``click'', or counting event. Voltage noise of amplitude $\sigma_V$ is rejected when $\sigma_V \ll V_\mathrm{disc} \ll V_0 - \sigma_V$. In practice, voltage noise has a negligible effect on counting-mode data whenever reasonable care is taken with the electronics: all signals carried by short coaxial cables, all circuits mounted in sealed metal cases, and all electronics powered by low-noise power supplies. \\

In counting mode, at most one click can be registered per time interval $\tdet$. If two particles are detected within the same time interval, the detector simply generates a larger voltage pulse within the same time interval. Only one click is generated, even though two particles have arrived. One could imagine setting another, higher discriminator voltage to account for two-particle arrivals. Unfortunately, this tactic fails because the amplitude of the detector pulses varies randomly from pulse to pulse over a wide range - as large as a factor of two. A second discriminator would frequently mis-identify single arrivals as doubles and vice versa. \\

Whenever the counting-mode detection rate becomes large compared to $1/\tdet$, we must account for the possibility of two-particle detection events being counted as a single click. The particle arrivals are uncorrelated within the time interval $\tdet$, so the event distribution is Poissonian. Writing $\alpha$ for the average number of detection events within $\tdet$, we see that the probability of $k$ events is given by
\begin{equation}
P_k(\alpha) = \frac{e^{-\alpha} \alpha^k}{k!}
\end{equation}
We \emph{fail} to observe a click in the interval $\tdet$ only if the number of successful detection events is zero, so the probability of observing a click is
\begin{equation}
P_\mathrm{click} = 1 - P_0(\alpha) = 1 - e^{-\alpha}
\end{equation}
We notice that $P_\mathrm{click} \approx \alpha$ for $\alpha \ll 1$ and that $P_\mathrm{click} \rightarrow 1$ as $\alpha \rightarrow \infty$. The average number of events is then
\begin{equation}
\alpha = -\ln (1 - P_\mathrm{click}) \label{clickrate}
\end{equation}
Even for $P_\mathrm{click} = 0.5$, we already have $\alpha = 0.69$, so that the approximation $P_\mathrm{click} \approx \alpha$ is wrong by 40\%.\\ \\

Counting detection becomes unreliable as $P_\mathrm{click}$ approaches $1$, since the error $\sigma_\alpha$ in our estimate of $\alpha$ becomes much larger than the error in the observed click probability $\sigma_{P_\mathrm{click}}$. Basic error propagation shows that \cite{Taylor-error-analysis-BOOK}
\begin{equation}
\sigma_\alpha = \frac{1}{1 - P_\mathrm{click}} \sigma_{P_\mathrm{click}} \label{clickerror}
\end{equation}
so that, for instance, the estimated error in $\alpha$ is ten times the error in $P_\mathrm{click}$ for $P_\mathrm{click} = 0.9$. Since it is difficult to obtain a reliable estimate for $\sigma_{P_\mathrm{click}}$ without a long data series, we should be cautious about applying Eqs. (\ref{clickrate}) and (\ref{clickerror}) for values of $P_\mathrm{click}$ close to 1. \\

Proportional-mode detection avoids the saturation effect of counting mode, so it becomes useful when high detection rates are required. If one wishes to measure the rate of some process over many orders of magnitude, as in experiments measuring total photoelectron or photoion yield, proportional mode is almost mandatory: in counting mode, the total measurement time becomes impractically long at the lowest rates. However, in proportional mode, it becomes difficult to distinguish between signal fluctuations and noise, and the details of the voltage waveform become important. Since counting mode is preferred for accurate measurements, we do not discuss the issues of proportional mode further in this review. \\

\subsection{Data accumulation and long-term noise}

For high-accuracy measurements, one almost always accumulates data for a time that far exceeds the interval between laser pulses. Accumulating data over many laser pulses, or ``shots'', improves the signal-to-noise ratio of the measured particle distribution and thus our ability to estimate the ``true'' distribution. Estimating the measurement uncertainty therefore requires statistical analysis of the data accumulation process. \\

Real experimental data is discrete: for a given particle arrival, the particle momentum is determined to lie within a finite volume element of momentum space. In other words, the momentum component along $x$ is measured as lying in the interval $(p_\mathrm{det,x}, p_\mathrm{det,x}+\Delta_\mathrm{det,x})$, and similarly for the momenta along $y$ and $z$. For simplicity, we denote this interval as $(\pdet,\pdet+\deldet)$. The detector resolution often varies with momentum, so in general $\deldet = \deldet(\pdet)$ is a function of momentum. Upon the $j$th laser shot, we obtain the data $d_j(\pdet)$, where $d_j(\pdet)$ is the (integer) number of particles that is detected in the momentum interval $(\pdet,\pdet+\deldet)$. \\

By accumulating data over $j = 1,\ldots N$ laser shots, we obtain $D_j(\pdet) = \sum_{j=1}^N d_j(\pdet)$, where $D_j(\pdet)$ is the total (integer) number of particles in the momentum interval. We can therefore estimate the average detection rate within a given momentum interval as $\drate(\pdet) = D_j(\pdet)/N$. If every laser shot is identical, the random error in determining $\drate$ can be estimated as \cite{Taylor-error-analysis-BOOK}
\beq
\sigma_d(\tau) \approx \frac{\sigma_D}{\sqrt{N}} = \frac{\sigma_D}{\sqrt{\tau \nu_\mathrm{rep}}} \label{shot-noise}
\eeq
where $\sigma_D$ is the random error in determining $D_j$ on each laser shot, $\tau$ is the total data accumulation time, or ``integration time'' and $\nu_\mathrm{rep}$ is the repetition frequency of the laser pulses. Eq. (\ref{shot-noise}) is closely related to the well-known ``shot noise limit'', and becomes identical to it if $\sigma_D \approx \drate^{1/2}$, i.e., when the $D_j$ follow a Poissonian distribution. \\

Regrettably, the conditions of the laser shots rapidly cease to be identical in real experiments. It is impractical to continuously monitor every relevant parameter and model its effects on the measurement result. Long-term drift of the experimental parameters begins to affect precision measurement results on the few-second timescale. Hence, the random error of the average value $\drate(\pdet)$ rises above the shot-noise value given by Eq. (\ref{shot-noise}). The uncertainties of measurements with integration times $\gtrsim 50$ s are far above the shot-noise value. These considerations also apply to comparisons between measurements taken at different times, e.g., when comparing electron spectra obtained at different laser intensities. Since the laser parameters continuously drift during the entire series of measurements, the random error in, e.g., the difference between two measurements can easily be much larger than the random error in each measurement. \\

\subsection{Uncertainty in a single measurement - Allan deviation}

We seek to quantify how the random error in the accumulated data, $\sigma_d(\tau)$, depends on the total integration time $\tau$. The Allan deviation is a robust statistical tool for this purpose that was originally developed for characterising clocks and frequency standards \cite{Allan-deviation-CLASSIC, Barnes-frequency-stability-allan-variance-REV} but is now used in many areas of metrology and precision measurement \cite{Allan-variance-general, Witt-allan-variance-tutorial}. The Allan deviation provides an estimate of $\sigma_d(\tau)$ through the formula
\beq
\sigma_d^2(\tau) = \frac{1}{2(m-1)} \sum_{j=1}{m} \left( \drate_{j+1}(\tau) - \drate_j(\tau) \right)^2 \label{allan}
\eeq
where $\drate_j$ is the average value measured between the time $j \tau$ and the time $j (\tau + 1)$. There are several refinements of the Allan deviation, e.g., the modified Allan deviation and the overlapping Allan deviation: see \cite{Pullen-Kielpinski-H-ionisation-THESIS} and references therein for a more complete discussion. Free software for computing the Allan deviation and its variants is widely available, e.g., as a MATLAB routine \cite{Hopcroft-allan-variance-matlab}. \\

The Allan deviation gives us a simple experimental prescription for estimating the random error $\sigma_d(\tau)$ for any proposed integration time $\tau$. We should take many repeated measurements of $\drate$, with each measurement being integrated over the smallest practical time interval $\tmin$, and the total time series of measurements spanning a time $\tmax$. Ideally, we should measure $D$ on every laser shot and record the time series for a total time $\tmax$ that is far longer than any experimentally relevant integration time. Then, applying the Allan deviation formula Eq. (\ref{allan}) to the time-series data, we obtain an estimate of the random error for any integration time $\tau \lesssim \tmax$. \\

\begin{figure}[thbp]
\begin{center}
\includegraphics[width=\columnwidth/2]{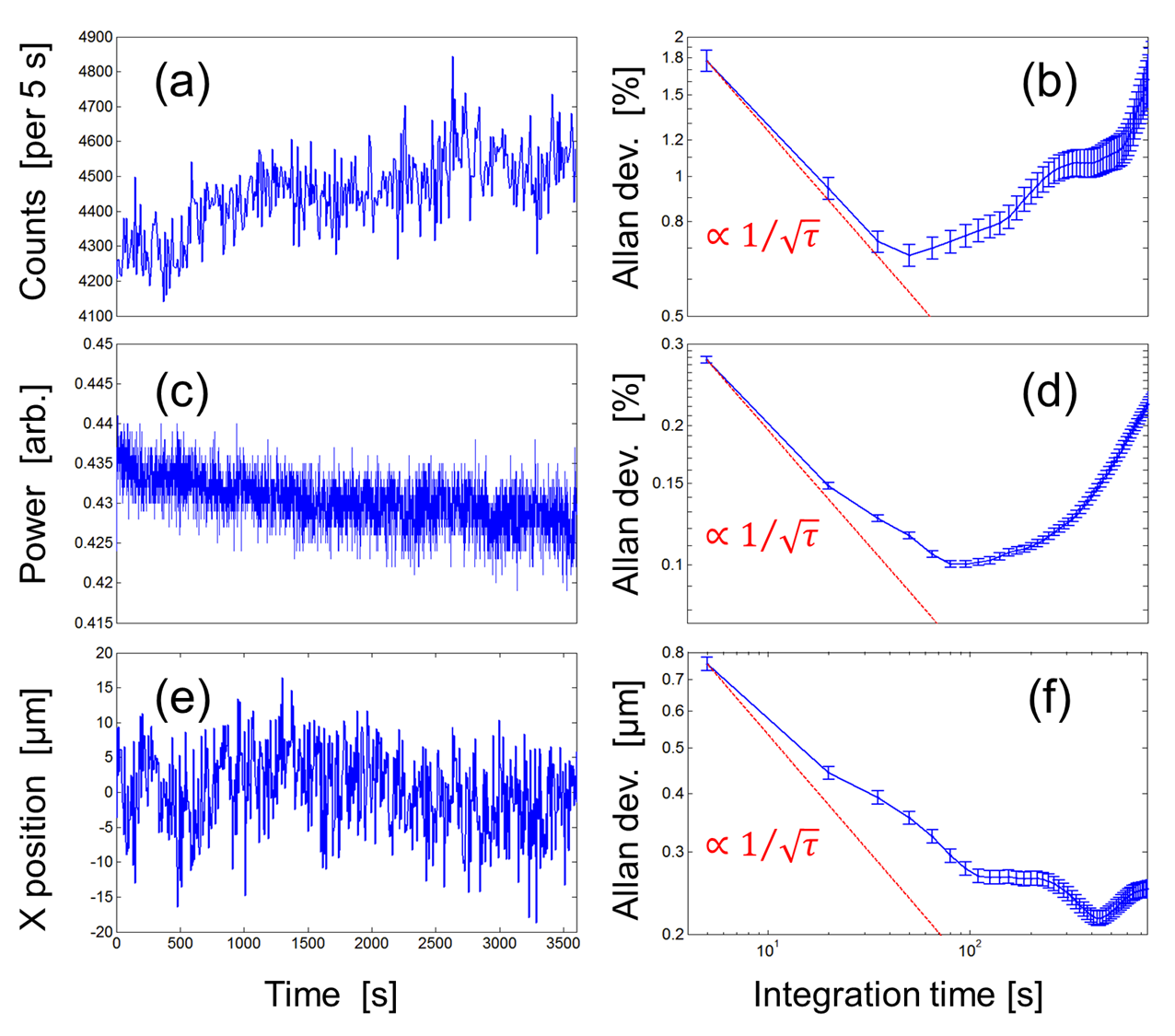}
\caption{Estimating random measurement error using the Allan deviation. The left column shows the measured time-series data; the right column shows the corresponding Allan deviation. Top row: photoelectron count rate; middle row: laser power; bottom row: laser position along atomic beam axis. Dashed lines in Allan deviation plots show the $1/\sqrt{\tau}$ limit set by Eq. (\ref{shot-noise}). Comparison of the Allan deviations shows that instability of the laser power may limit the useful integration time, while instability of the laser position probably does not.}
\label{fig-allan}
\end{center}
\end{figure}

Figure \ref{fig-allan} illustrates the use of the Allan deviation in a strong-field photoionisation experiment. Panel (a) shows a time series of measurements of photoelectron counts taken at the AASF. Each point in the time series represents a measurement accumulated over an interval $\tmin = 5$ s, while the total time series lasts for $\tmax = 3600$ s. Panel (b) shows the corresponding error estimate. For short integration times $\tau \lesssim 40$ s, we see that the Allan deviation is proportional to $1/\sqrt{\tau}$, as predicted by the Eq. (\ref{shot-noise}). On this timescale, drift of the experimental parameters has not yet set in. However, for integration times $\tau \sim$ 60 s, the Allan deviation no longer decreases owing to drift. In planning an experiment, we would say that there is probably \emph{no point} in integrating each measurement for longer than 40 -- 50 s, as we will not improve our measurement precision by integrating for a longer time. Finally, for integration times $\tau \gtrsim 100$ s, the precision of the measurement actually grows \emph{worse}. By no means should we extend the integration time beyond 100 s. This kind of reasoning has been employed extensively and successfully in planning precision measurement experiments at the AASF \cite{Pullen-Kielpinski-few-cycle-H-ionisation, Pullen-Kielpinski-H-ionisation-THESIS, Laban-Sang-zeptosecond-XUV-interferometer, Wallace-Kielpinski-H-ionisation-CEP-effects}. \\

The Allan deviation also provides clues as to the experimental sources of long-term measurement error. Figure \ref{fig-allan}(c) shows a time series of measurements for a key experimental parameter: the power of the AASF laser system, as detected by a standard photodiode. These measurements can be converted to an Allan deviation through Eq. (\ref{allan}) in the same way as the photoelectron counts. The results are shown in Figure \ref{fig-allan}(d). We see that the Allan deviation of the laser power starts to increase for integration times $\tau \sim$ 100, the same timescale as for the Allan deviation of the photoelectron counts in Figure \ref{fig-allan}(b). While this similarity does not prove that the laser power noise definitely causes the photoelectron count noise, it indicates that there may be an effect worth investigating. On the other hand, Figure \ref{fig-allan}(e,f) show the time series and Allan deviation for the position of the laser beam along the propagation direction of the atomic beam. The Allan deviation continues to decrease for several hundred seconds, so the instability of the laser beam position does not cause the observed instability in the photoelectron count at the $\sim 100$ s timescale. \\

\subsection{Uncertainty of measurement comparisons}

For comparisons of measurements taken at different times, we have an additional source of random error due to the drift between measurements. The Allan deviation only quantifies the random error for an individual measurement, in which data is collected during the single-measurement integration time $\tau$. However, if two measurements are separated by a ``dead time'' $T \gg \tau$, the experimental parameters will still continue to drift during the dead time, and therefore the measurement results will also deviate from each other by an amount that exceeds their individual uncertainties. \\

One often considers a measurement as a member of a series, for instance, in determining the intensity dependence of photoelectron yield. A conservative error estimate will assume that the dead time between measurements is equal to the entire acquisition time of the series, $T_\mathrm{ser}$. (Ideally, the measurements are taken in a random order unknown to the experimenters.) The random error of each measurement in the series can then be estimated as
\beq
\sigma_{d,ser} = \sqrt{\sigma^2_d(\tau) + \sigma^2_{d,\mathrm{dead}}(T_\mathrm{ser})}
\eeq
where $\sigma_{d,\mathrm{dead}}(T_\mathrm{ser})$ is the additional random error arising from the dead-time drift over the time $T_\mathrm{ser}$. \\

The dead-time error $\sigma_{d,\mathrm{dead}}(T)$ can be quantified from time-series measurements by means of the power spectral density (PSD) of the noise. Just as an optical spectrometer measures the power spectrum of the (time-dependent) electric field, the PSD quantifies the power spectrum of the noise signal \cite{Paschotta-laser-encyclopedia}. To this end, suppose we obtain a reference time series of $m$ measurements of $\drate$, spanning a total time $\tmax$. The noise PSD can be estimated as \cite{Shanmugan-Breipohl-random-signal-analysis-BOOK}
\beq
S_{dd}(\nu) = \frac{1}{m} \left| \sum_{j=1}^{m} \drate_j e^{-2 \pi i j \nu} \right|^2
\eeq
which is just the squared magnitude of the discrete Fourier transform of the series $\{\drate_j\}$. The variance between measurements of $\drate$ taken at a dead time separation of $T$ is then given by
\beq
\sigma^2_{d,\mathrm{dead}}(T) = \int_{1/T}^{1/\tmin} \!d\nu\: S_{dd}(\nu)
\eeq
where $\tmin$ is the time interval for each measurement of $\drate$. \\

The estimated PSD is often found to be quite noisy for time-series measurements of reasonable duration. However, the PSD of measurements on physical systems almost always follows the form
\beq
S(\nu) = \left\{ \begin{array}{lll} A/\nu^\alpha & \mbox{if } \nu \leq \nu_c & \mbox{``flicker noise''} \\ B & \mbox{if } \nu > \nu_c & \mbox{``white noise''} \end{array} \right.
\eeq
where $A$, $B$, $\alpha$, and $\nu_c$ are constants. By fitting this model to the estimated PSD, one can avoid anomalies in the estimation of $\sigma_{d,\mathrm{dead}}(T)$. \\

\section{Principles for predicting strong-field measurement results}
\label{sec-compare}

Undertaking an accurate theory-experiment comparison in strong-field ionisation requires careful modelling of the experimental conditions as well as an accurate atomic theory of the strong-field interaction. With generally available experimental techniques, the measured electron distribution cannot be expected to reproduce the output of single-atom simulations.\footnote{There are prospects for improving this situation. A recent experiment \cite{Kahra-Schaetz-single-ion-strong-field} observed the dissociation dynamics of an isolated molecular ion in a UV laser field of $\sim 10^{11} \:\mbox{W}/\mbox{cm}^2$ intensity.} Instead, one measures the distribution that originates from many atoms, each of which is subjected to somewhat different conditions. We therefore need to predict the measured distribution by (1) integrating the single-atom theoretical predictions over variations in the interaction and (2) propagating the integrated distributions through a model of the detector system. The resulting prediction can be expected to match the measurement, but is often quite different from the single-atom theoretical prediction. \\

\subsection{Theoretical predictions of strong-field ionisation dynamics}

First, we remind the reader of what, exactly, constitutes a theoretical single-atom prediction for strong-field ionisation. At nonrelativistic intensities ($\lesssim 10^{16} \:\mbox{W}/\mbox{cm}^2$) strong-field dynamics in atomic hydrogen is governed by the time-dependent Schr\"odinger equation (TDSE) for the electron
\beq
i \hbar \frac{\partial}{\partial t} \psi(\myvec{r},t) = \left[ \frac{p^2}{2 m} - e V(\myvec{r}) - e \myvec{E}(t) \cdot \myvec{r} \right] \psi
\eeq
where $\psi = \psi(\myvec{r},t)$ is the electron wavefunction, $V(\myvec{r}) = 1/(4 \pi \epsilon_0 |\myvec{r}|)$ is the Coulomb potential, $\myvec{E}(t)$ is the laser electric field, and $\myvec{p} = -i \hbar \myvec{\nabla}$ is the momentum operator. \\

For atomic H, the exact TDSE can be solved numerically for 800 nm laser wavelength and intensities of $10^{14}$ to $10^{15} \:\mbox{W}/\mbox{cm}^2$ by expending substantial time on medium-size computer clusters. The majority of strong-field ionisation experiments fall in this regime. The difficulty of the computation scales rapidly with $U_p$, so increasing the wavelength or intensity can radically increase computation time. For any atom or molecule with more than one electron or more than one nucleus, the computational problem of the exact TDSE in this regime becomes too difficult, and approximations must be used. As discussed in the Introduction, a major advantage of experiments with atomic H is the opportunity to test these approximations. \\

The most sophisticated electron detection systems are set up to report the measured momentum $\pdet$ of the outgoing electrons. Almost all detectors can be considered as measuring some function of the momentum distribution, e.g., the kinetic energy. The laser-atom interaction is long over by the time the electrons are detected, so we can let $t \rightarrow \infty$. The electron momentum distribution is proportional to the momentum-resolved ionisation probability,
\ba
\Pi(\myvec{p}) &=& \lim_{t \rightarrow \infty} \left| \Braket{ \myvec{p} | \Psi(\myvec{r},t) } \right|^2 \\
&=& \lim_{t \rightarrow \infty} \left| \int \! d\myvec{r} \: e^{i \myvec{p} \cdot \myvec{r}/\hbar} \Psi(\myvec{r},t) \right|^2
\ea
Various numerical techniques exist for accurately dealing with the $t \rightarrow \infty$ limit, so a momentum distribution calculated from an accurate numerical TDSE wavefunction is also accurate. \\

From the experimental point of view, it is crucial to realise that $\Pi(\myvec{p})$ is evaluated for a completely specified laser electric field waveform $\myvec{E}(t)$. For the sake of convenience, numerical calculations may use a laser pulse shape that is quite different from the experimentally observed pulse shape. Experimentalists usually characterise the laser field by its intensity, carrier-envelope phase, and pulse shape, with the characterisation of pulse shape presenting special difficulty for few-cycle pulses. All these quantities can fluctuate in time and may also vary spatially depending on the focusing conditions. So far, it appears that the exact pulse shape is relevant for the high-energy plateau electrons of energy $\gtrsim 5 U_p$ (see, e.g., \cite{Chen-Lin-CEP-ATI-waveform-retrieval}), but is not very important at lower energies \cite{Pullen-Kielpinski-H-ionisation-THESIS, Pullen-Kielpinski-peak-intensity-calibration}. However, both the intensity \cite{Pullen-Kielpinski-few-cycle-H-ionisation, Pullen-Kielpinski-peak-intensity-calibration} and the carrier-envelope phase \cite{Wallace-Kielpinski-H-ionisation-CEP-effects} are critical parameters over the entire electron energy range. \\

\subsection{Laser beam geometry}

In many experiments, the laser beam can be approximated as a Gaussian beam (\cite{Saleh-Teich-fundamentals-photonics-BOOK}, ch. 3). For an ideal Gaussian beam propagating along $z$ and focused at $z = 0$, we have
\ba
I(\rho,z) &=& I_0 \left( \frac{w_0}{w(z)} \right)^2 e^{-2 \rho^2/w^2(z)} \label{gaussintens} \\
w(z) &\equiv& w_0 \sqrt{1 + \left( \frac{z}{z_R} \right)^2} \label{w(z)}
\ea
Here $I_0$ is the peak intensity at the centre of the laser focus, $w_0$ is the $1/e^2$ intensity radius of the laser spot at the focus, $\rho$ is the distance from the $z$ axis, $z_R = \pi w_0^2/\lambda$ is the Rayleigh range, and $\lambda$ is the laser wavelength. The parameter $w(z)$ measures the $1/e^2$ spot radius along the propagation axis. \\

The laser beam geometry also affects the carrier-envelope phase. For a Gaussian beam, the carrier-envelope phase of the pulse evolves according to the Gouy phase
\ba
\zeta(z) = \tan^{-1} \frac{z}{z_R} \label{gouy}
\ea
The effects of Gouy phase have been directly measured in experiments on strong-field ionisation \cite{Lindner-Krausz-few-cycle-Gouy-phase} and high-harmonic generation \cite{Laban-Sang-zeptosecond-XUV-interferometer} with noble-gas targets. \\

The ideal Gaussian beam model may be insufficient when the laser-target interaction region has a significant size along the laser propagation direction. Deviations from an ideal Gaussian beam are especially common for few-cycle pulses, since the spatial mode is easily distorted by hollow-fibre compression and also by misalignment of the reflective focusing optics. While we use the ideal Gaussian beam model in the remainder of this review, Appendix \ref{sec-aberrations} describes an extended model that is generally adequate to describe non-Gaussian beams encountered in strong-field ionisation experiments. The calculations below can be easily modified to take account of the extended model. \\

A few experiments in strong-field physics employ special optics to obtain a more uniform intensity distribution than permitted by a Gaussian beam \cite{Boutu-Carre-flat-top-beam, Fu-Chin-flat-top-beam}. These techniques show great promise for improving the precision measurement of strong-field ionisation. However, they have not yet been applied to this problem and fall outside the scope of this review. \\

\subsection{Target interaction geometry}

\begin{figure}[thbp]
\begin{center}
\includegraphics[width=\columnwidth]{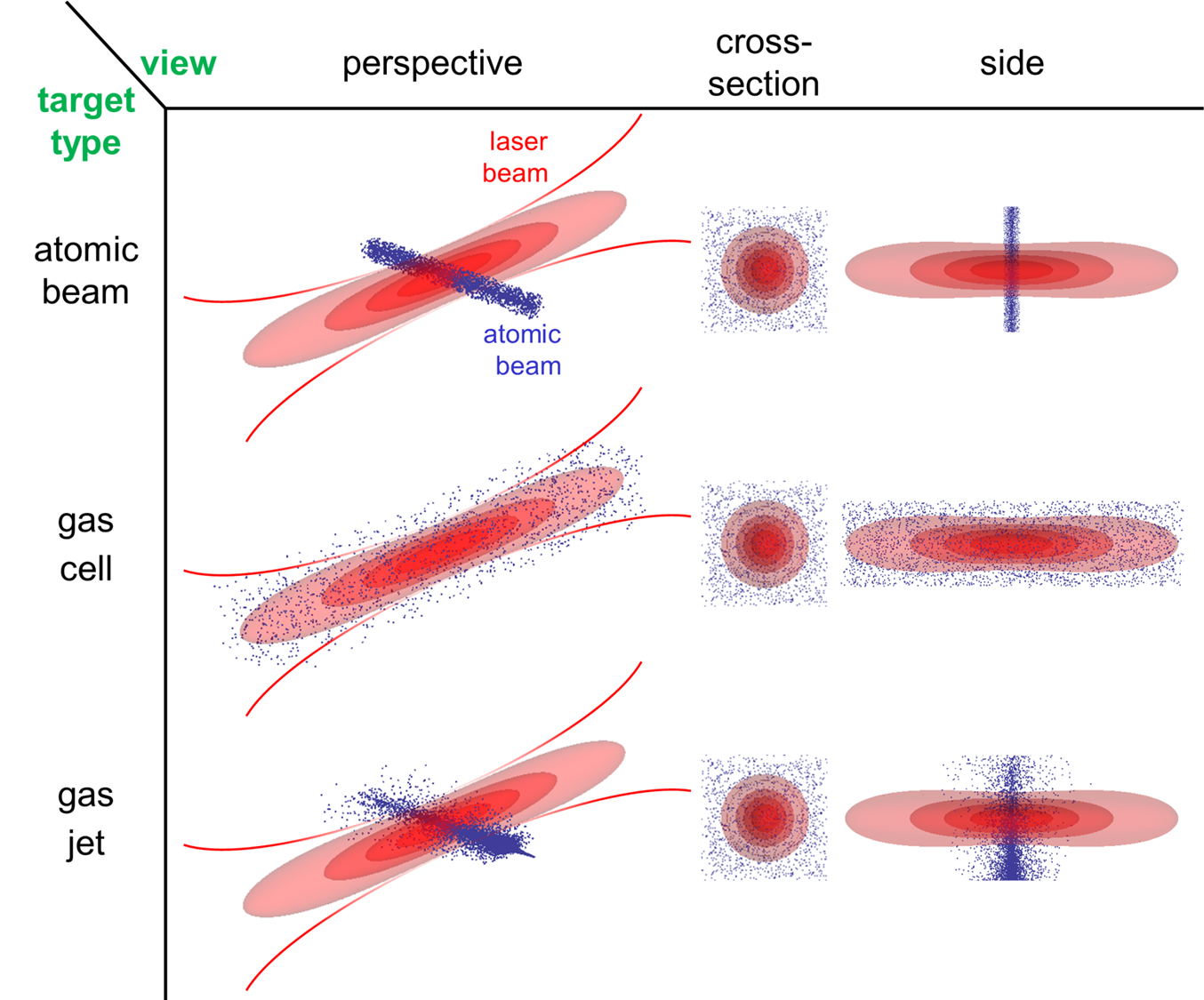}
\caption{Laser-target interaction geometries for strong-field ionisation. Each row depicts the geometry for one of the three types of gas targets: atomic beam, gas cell, and gas jet. Left column: perspective view of the interaction geometry. Middle column: cross-section in the laser focal plane ($z=0$). For all three target types, the atomic distribution is independent of the transverse position in the laser beam. Right column: side view perpendicular to both the laser beam and the gas target. Red shading: laser intensity contours; blue dots: atomic distributions; red lines: laser beam edges.}
\label{InteractionGeometries}
\end{center}
\end{figure}

The interaction between the laser and the gas target is determined by the number density of atoms in the target, $n_\mathrm{at}$. Gas targets naturally fall into three classes: (1) jet targets, in which the gas is emitted from a nozzle, (2) molecular beams, in which a gas jet propagates freely through vacuum and is collimated by apertures, and (3) gas cells, in which the gas uniformly fills a region. Jet targets typically exhibit a smoothly varying, nonuniform density in the interaction region. For molecular beams and gas cells, the density is uniform within the interaction volume, and zero outside the volume. The laser-target interaction geometries for these target types are depicted in Figure \ref{InteractionGeometries}. For each type of target, the spot size $w(z)$ of the laser beam typically remains much smaller than the length scale of the molecular density variations. \\

Although gas jets are the most common type of target, the laser-target interaction geometry is poorly defined. In a jet target, atoms are fed into a nozzle from a constant-pressure gas source. The atoms spread out as they emerge from the nozzle, so even if the density is uniform within the nozzle, it quickly assumes a smooth, nonuniform distribution. An effusive jet, with a wide nozzle and low backing pressure, has a $\cos \theta$ angular distribution, where $\theta$ is measured from the normal to the nozzle aperture. On the other hand, a narrow nozzle at high pressure will create a supersonic expansion, which can have a very narrow angular distribution. \\

Atomic beam targets create a sharply defined interaction region by removing the ``soft edges'' of the gas jet itself. The atomic beam can be made thin relative to the laser Rayleigh range $z_R$, but wide relative to the laser spot size $w(z)$. Within the beam, the density is uniform, but outside the beam, it is zero. The atomic beam is formed when a gas jet passes through a high-vacuum flight path containing two thin apertures. The background pressure is kept so low that atoms in the beam do not scatter from background gas in the flight path. Each atom therefore flies along an independent, straight-line trajectory. When the beam encounters an aperture, each atom inside the open part of the aperture passes through undisturbed, but all other atoms are lost from the beam. The diameters of the apertures are typically much smaller than the length of the flight path, so to pass both apertures, a atom must fly straight along the line between them. All atoms emerging from the second aperture are flying parallel to each other, so the density profile of the beam is maintained between the second aperture and the interaction region. \\

The gas cell is common in experiments with high-energy pulses and/or high-pressure targets. The cell is simply a sealed chamber connected to a constant-pressure gas source. The gas may even just originate in outgassing from the walls of the vacuum chamber. Often the laser beam enters and exits through optical windows. In this case the ionisation rate is guaranteed to be negligible at the entrance and exit; otherwise the windows would immediately be optically damaged. The atomic number density is constant within the cell. Alternatively, the cell may sit in a larger vacuum chamber and the laser beam may enter and exit through small apertures. The number density will then become slightly nonuniform near the apertures. \\

\subsection{Carrier-envelope phase averaging}
\label{sec-cep-avg}

Theoretical calculations assume a completely characterised electric field waveform, implying a fixed value of carrier-envelope phase (CEP). Experimentally, the CEP of consecutive pulses from a laser varies randomly over the entire $2\pi$ range unless some CEP stabilisation technique is employed. The most common technique measures the CEP with a $f-2f$ interferometer and feeds back the resulting error signal to the laser system, stabilising the CEP \cite{Baltuska-Krausz-phase-stable-amplifier}. Another active stabilisation technique uses the CEP dependence of strong-field photoelectron spectra to obtain the error signal \cite{Adolph-Paulus-stereo-ATI-CEP-locking}. One can also use optical parametric amplification to obtain high-energy pulses with passive CEP stability \cite{Vozzi-Cerullo-CEP-stable-midIR-amplifier}. Nevertheless, it appears challenging to obtain CEP noise below 300 mrad RMS at the laser-atom interaction region, even over timescales of a few seconds \cite{Wittmann-Kienberger-single-shot-CEO-meast}. The influence of changing air currents over a several meter beam path, say between the $f-2f$ interferometer and the interaction region, is sufficient to substantially degrade CEP stability. \\

\begin{figure}[thbp]
\begin{center}
\includegraphics[width=\columnwidth/2]{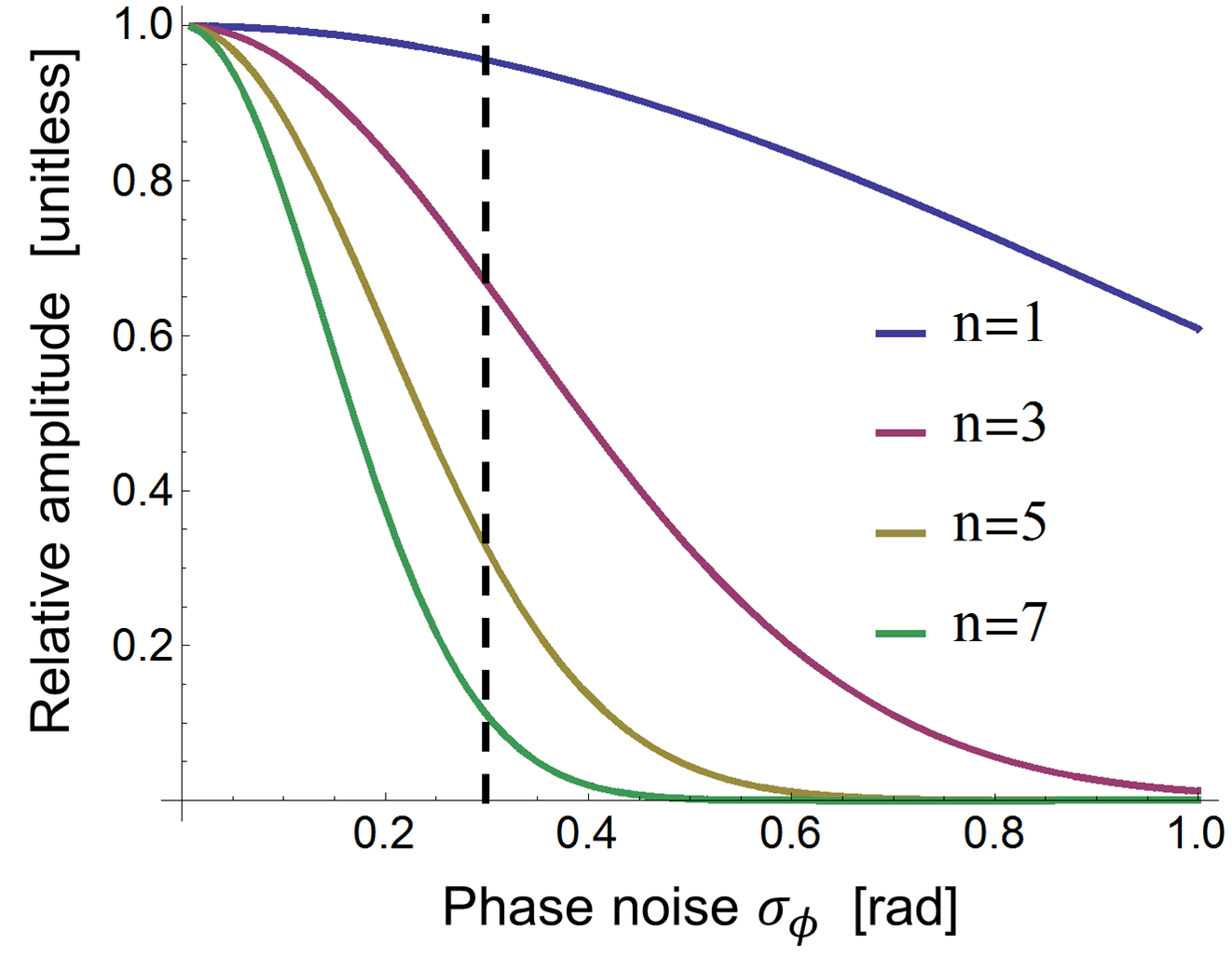}
\caption{Reduction of CEP effects due to phase noise. The amplitude reduction factors $\kappa_n(\scep)$ from Eq. (\ref{kappa-n}) are plotted for the first few values of $n$. While the lowest-order CEP effect is quite robust to phase noise, the details of CEP variation (higher $n$) are easily lost for currently accessible levels of phase noise in experiments (vertical dashed line). }
\label{phasenoise}
\end{center}
\end{figure}

To compare experiment with theory, we must therefore average the theoretical predictions $\Pi(\myvec{p},\cep)$ over the measured distribution of CEP noise. Of course, for a CEP-averaged measurement we must average over a uniform CEP distribution $0 \leq \cep < 2 \pi$. We can expand the CEP-dependent momentum distribution in a Fourier series:
\ba
\Pi(\myvec{p},\cep) = \Pi_0(\myvec{p}) + \sum_n \Pi_n(\myvec{p}) \sin n(\cep + \phi_n(\myvec{p}))
\ea
where the coefficients $a_2, a_4, \ldots$ are generally much smaller than $a_1, a_3, \ldots$ \cite{Roudnev-Esry-general-CEP-theory}. For a Gaussian distribution of CEP noise with standard deviation $\scep$, we find that the electron distribution becomes
\ba
\Pi_\mathrm{av}(\myvec{p},\cep) &=& \int_0^{2\pi} \! d\phi' \: e^{-\phi'^2/(2 \scep^2)} \Pi(\myvec{p},\phi') \\
&=& \Pi_0(\myvec{p}) + \sum_n \kappa_n(\scep) \Pi_n(\myvec{p}) \sin (\cep + \phi_n(\myvec{p})) \\
\mbox{with } \kappa_n(\scep) &\equiv& e^{-n^2 \scep^2} \:\mbox{Re}\! \left[ \frac {\erf \left[ (\pi + i n \scep^2)/(\sqrt{2} \scep) \right] }{\erf \left[ \pi/(\sqrt{2} \scep) \right]} \right] \label{kappa-n}
\ea
While the phase offsets $\phi_n$ of the CEP modulation terms are not affected by the noise, the amplitudes are reduced by a factor $\kappa_n(\scep)$ that depends sharply on the CEP modulation frequency $n$. The $\kappa_n(\scep)$ are plotted for the first few values of $n$ in Figure \ref{phasenoise}. \\

These CEP noise considerations show that precision measurement has a key role to play in revealing the attosecond dynamics of ionisation. For current experiments, with their phase noise of $\sim 300$ mrad at the interaction region, it appears that only the CEP oscillations with $n = 1$ (period $2\pi$) and $n = 3$ (period $2 \pi/3$) are readily accessible. The $n = 3$ oscillation corresponds to a time period of $\sim 1$ fs at the commonly used 800 nm laser wavelength. Attosecond timing requires us to push to $n \geq 5$, for which the observable CEP effects are an order of magnitude smaller than the generally observed $n = 1$ oscillation effects. Precision measurement techniques become crucial for reliable observations of such small signals. \\

A particularly common CEP-resolved measurement uses two TOFs on opposite sides of the interaction region to detect high-energy electrons emitted parallel and antiparallel to the laser polarisation direction \cite{Paulus-DeSilvestri-CEO-depdt-ionization}. This setup is often used as a means of retrieving the CEP \cite{Paulus-Krausz-phase-depdt-electron-emission} and providing a CEP stabilisation signal to the laser system \cite{Adolph-Paulus-stereo-ATI-CEP-locking}. The two TOFs nominally report the electron spectra $\Pi_U(E_e,\cep)$ and $\Pi_D(E_e,\cep)$, where $E_e$ is the electron energy. Neglecting detector imperfections, we have $\Pi_D(E_e,\cep) = \Pi_U(E_e,\cep+\pi)$ by symmetry. The CEP effects are often parametrised in terms of the asymmetry:
\ba
A(E_e,\cep) &\equiv& \frac{\Pi_U(E_e,\cep) - \Pi_D(E_e,\cep)}{\Pi_U(E_e,\cep) + \Pi_D(E_e,\cep)}
\ea
For all but the shortest pulses ($\lesssim 4$ fs), the CEP variations of the sum $\Pi_U + \Pi_D$ are much smaller than the difference $\Pi_U - \Pi_D$. Hence, the amplitude of the asymmetry oscillations at period $2\pi/n$ is reduced by the same factor $\kappa_n$ given in Eq. (\ref{kappa-n}). For the very shortest pulses, $\Pi_U + \Pi_D$ depends nontrivially on the CEP and there is no simple closed-form expression for the CEP noise effects on the asymmetry.  \\

\subsection{Focal volume integration}
\label{sec-fvi}

The photoelectron yield of strong-field ionisation depends strongly on the intensity, so spatial variations of the intensity within the laser-atom interaction region are a major effect in theory-experiment comparisons. The number density of atoms $n_\mathrm{at} = n_\mathrm{at}(\vec{r})$ may also vary within the interaction region. However, as mentioned above, theoretical calculations provide the momentum distribution $\Pi(\myvec{p})$ at a constant peak intensity $I$. The resulting disparity between experiment and theory has been long recognised \cite{Augst-Chin-barrier-suppression-ionisation} and was qualitatively discussed for the case of hydrogen in \cite{Paulus-Walther-hydrogen-ATI-expt}. To include this ``focal volume integration'' (FVI) effect, we must integrate the momentum distribution over the distribution of laser peak intensities, obtaining
\ba
S(\myvec{p}) = \int \! d\myvec{r} \: n_\mathrm{at}(\myvec{r}) \: \Pi \!\left(\myvec{p};I(\myvec{r})\right) \label{fvi-eq}
\ea
where $I(\myvec{r}) = I(\rho,z)$ is given by the Gaussian beam intensity, Eq. (\ref{gaussintens}). The FVI distribution $S(\myvec{p})$ gives the momentum distribution of all electrons (or ions) created in the interaction region. \\

FVI can also induce additional CEP averaging through the Gouy phase. If the interaction region is significantly extended along the laser propagation direction, the CEP will vary within the interaction region. We have
\ba
S(\myvec{p},\cepzero) = \int \! d\myvec{r} \: n_\mathrm{at}(\myvec{r}) \: \Pi \!\left(\myvec{p};I(\myvec{r}),\cep(z)\right) \label{fvi-gouy-eq}
\ea
where $\cepzero$ is the CEP at $z=0$ and $\cep(z) = \cepzero + \tan^{-1} (z/z_R)$ according to Eq. (\ref{gouy}). \\

Atomic beam experiments permit simple and reliable evaluation of FVI effects. The atomic number density is constant inside the beam, $n_\mathrm{at} = n_0$, and zero outside. We set the atomic beam diameter $d_\mathrm{beam}$ so that $w_0 \ll d_\mathrm{beam} \ll z_R$. The laser focus intersects the centre of the atomic beam at right angles. The interaction region is approximately a cylinder with diameter $w_0$ and length $d_\mathrm{beam}$. The faces of the cylinder are flat, since $w_0 \ll d_\mathrm{beam}$. The laser spot size does not change appreciably in the interaction region since $d_\mathrm{beam} \ll z_R$. The FVI equation (\ref{fvi-eq}) takes on the ''2D'' form
\beq
S_\mathrm{beam}(\myvec{p}) = 2 \pi d_\mathrm{beam} n_0 \int_0^\infty \! \rho \: d\rho \: \Pi \!\left(\myvec{p}; I_0 e^{-2 \rho^2/w_0^2}\right) \label{fvi-2D}
\eeq
If the laser focus is in the centre of the atomic beam, the Gouy phase shift over the interaction region can be estimated as $\Delta \phi_\mathrm{Gouy} \approx \pi d_\mathrm{beam}/z_R$. The effect on CEP-resolved measurements is negligible as long as $\Delta \phi_\mathrm{Gouy}$ is small compared to the CEP noise $\sigma_\phi \gtrsim 300$ mrad. In practice, one therefore requires $d_\mathrm{beam} < z_R/10$. \\

FVI can also be straightforward for gas cell experiments, since the atom number density can be assumed constant everywhere inside the cell. If the laser beam conforms to the ideal Gaussian model of Sec. \ref{sec-laser}, the FVI equation (\ref{fvi-eq}) simplifies to the ``3D'' form
\beq
S(\myvec{p}) = 2 \pi n_0 \int_{z_\mathrm{in}}^{z_\mathrm{out}} \! dz \int_0^\infty \! \rho \: d\rho \: \Pi \!\left(\myvec{p};I(\rho,z)\right) \label{fvi-3D}
\eeq
with $z_\mathrm{in,out}$ being the positions of the cell entrance and exit, measured relative to the position of the laser focus, and $I(\rho,z)$ given by Eq. (\ref{gaussintens}). In this case, one can approximate $z_\mathrm{in} \rightarrow -\infty$ and $z_\mathrm{out} \rightarrow \infty$, and the simplified FVI equation (\ref{fvi-3D}) should be extremely accurate. As mentioned above, gas cells that use pinholes for entrances and exits will experience density variations near the pinholes, degrading the accuracy of Eq. (\ref{fvi-3D}). \\

Even though the mathematics of FVI is simple for both gas cells and molecular beams, the data from molecular beams is easier to interpret than the data from gas cells. Features like the above-threshold ionisation plateau are already harder to see in molecular beam data than in theoretical predictions because of FVI over the 2D laser slice. Since gas cells incur additional FVI along the laser propagation direction, such features will be washed away further. In CEP-resolved measurements, the Gouy phase shift along the propagation direction also causes substantial CEP averaging:
\beq
S(\myvec{p},\cepzero) = 2 \pi n_0 \int_0^\infty \! \rho \: d\rho \: \int_{z_\mathrm{in}}^{z_\mathrm{out}} \! dz \: \Pi \!\left(\myvec{p};I(\rho,z),\cepzero+\zeta(z)\right) \label{fvi-gouy-3D}
\eeq
The more complex FVI for gas cells also requires a much more careful evaluation of the laser beam quality, which can easily depart from the ideal Gaussian model. If significant laser beam aberrations are present, one must use the more complex laser model of Appendix \ref{sec-aberrations} in conjunction with the full FVI equation (\ref{fvi-eq}) to evaluate the FVI. \\

There is no straightforward way to evaluate FVI for gas jets. As discussed above, a typical gas jet exhibits a smooth variation of $n_\mathrm{at}(\vec{r})$ that is not known \emph{a priori}. If the gas jet diameter becomes significant compared to $z_R$, the details of $n_\mathrm{at}(\vec{r})$ become important. Reliable FVI will then require a careful simulation of the gas dynamics or auxiliary experimental characterisation of the 3D number density. It will also be necessary to carefully characterise the laser beam using the model of Appendix \ref{sec-aberrations}. \\

To simplify FVI for a gas jet target, one can block the electrons (or ions) that originate outside a restricted interaction volume \cite{Hansch-VanWoerkom-slit-selection-interaction-region}. A slit is placed near the interaction region, between the interaction region and the detector. Suppose the electrons fly in a straight line to the detector, and only the particles that pass the aperture reach the detector. Working backward along the line of sight from the detector, we see that the slit defines a sharp slice of the interaction region. FVI becomes simplest for a long slit, with length perpendicular to the laser propagation direction and width $d_\mathrm{slit} \ll z_R$. In this case, the FVI reduces to the same form as for the molecular beam, Eq. (\ref{fvi-2D}), with $d_\mathrm{beam}$ replaced by $d_\mathrm{slit}$. \\

For ultimate accuracy, the atomic beam is probably preferable to the arrangement of the gas jet plus slit. In the latter case, one may worry about the effect of any stray electric or magnetic fields in the vicinity of the interaction region, which would break the assumption that the ionised particles travel in a straight line through the slit. The apertures in the atomic beam only need to block neutral particles, which are relatively unaffected by stray fields. The gas jet density should also be low enough that the ionised particles do not scatter from the neutral gas as they emerge; a properly operating atomic beam generally satisfies this condition, since the beam is thin and the density is in the molecular flow regime. However, we are not aware of any experiments that have rigorously compared the accuracy of the two arrangements. \\

\subsection{Detector system modeling}
\label{sec-detector-model}

Real electron detector systems never have perfect momentum resolution. In fact, no detector system measures momentum directly - the raw data consists of electronic pulses that register the arrival time and/or arrival position at a multichannel plate or channel electron multiplier. Knowledge of the geometry and electron-optical design of the detector system then permits one to estimate the properties of the momentum distribution. \\

In the most common electron detection system, the time-of-flight spectrometer (TOF), the photoelectrons leave the interaction region, propagate freely through a field-free space, and arrive at a microchannel plate. Dividing the field-free path length by the measured time of flight gives the velocity $v$ along the detection direction. The solid angle of electron collection is determined by the path length, the microchannel plate diameter, and sometimes, apertures in the flight path. \\

While TOF timing measurements can be made quite accurate, much more work is required to ensure that the inferred electron velocity is equally accurate. The length of a given electron trajectory is not simply equal to the physical path length and the electron velocity is not necessarily constant. Stray electric and magnetic fields can easily distort the electron trajectory. In many cases, electron optics are used near the interaction region to increase the solid angle of collection, and their fields also change the trajectory. The effects of these fields depend on the initial electron velocity and direction. Hence, the time of flight can deviate from the expected $1/v$ dependence and the solid angle $\Omega$ also depends on $v$. \\

The TOF is not a special case - we have used it as an example to illustrate similar systematic effects that are present in all electron detection systems. More generally, our detection system gives us some raw data, e.g., some time-dependent voltage readouts $V_j(t)$, which are subject to noise processes. Our task is to estimate the electron momentum distribution $S(\myvec{p})$ at the interaction region. The detection system may also have some auxiliary settings $W_k$ that we can change in order to get more information. The field of estimation theory deals with methods for performing this task. Fortunately, all detection systems share some common features that simplify the problem. \\

To translate the predicted electron distribution from the interaction region to the detector, we define the detector function $\kappa(\pdet|\myvec{p},\auxpars)$. The detector function is equal to the probability that a particle is observed at momentum $\pdet$, conditional on the particle having initial momentum $\myvec{p}$, and given the auxiliary detector parameters $\auxpars = V_1,\ldots,V_n$. (This notation conforms to the standard probability notation $P(A|B)$, read as ``the probability of observing $A$ conditional on the given facts $B$''.) The auxiliary parameters might include, e.g., the voltages applied to electron-optical elements. Once the detector function is known, we can predict the momentum distribution observed by the detector:
\beq
Y(\pdet;\auxpars) = \int \! d\myvec{p} \: \kappa(\pdet|\myvec{p},\auxpars) S(\myvec{p}) \label{momdet}
\eeq
where $S(\myvec{p})$ is the focal-volume-integrated momentum distribution. For detectors that do not resolve the full momentum distribution, the measured momentum $\pdet$ should be replaced with the measured property, e.g., the electron energy in the case of a TOF detector. \\

Detector systems are generally cylindrically symmetric about an axis $\hat{\zdet}$, and the so-called energy-resolving detectors, such as TOFs or repellers, actually measure the momentum component $p_\zdet$ along the detector axis. (In the literature, the detector axis is usually labeled $\hat{z}$. Here we denote it by $\hat{\zdet}$ to avoid confusion with the laser optic axis $\hat{z}$.)  Ideally, the angle $\theta$ between $\myvec{p}$ and the $\hat{\zeta}$ axis should satisfy $\theta \ll 1$ for all the detected electrons. The energy $E$ can then be written as
\beq
E \approx \frac{p_\zdet^2}{2 m} \left( 1 + \frac{1}{2} \theta^2 + \ldots \right)
\eeq
If the range of $\theta$ is sufficiently small, we can neglect $\theta$ entirely and write $p_\zdet = \sqrt{2 m E}$. Writing $\phi$ for the azimuthal angle about the $\hat{\zdet}$ axis, the momentum distribution $\Pi(\myvec{p})$ can be rewritten as
\beq
\Pi(E,\theta,\phi) = \sqrt{\frac{m}{p_\zdet}} \Pi(\myvec{p})
\eeq
where the proportionality constant arises from the change of variables. For a general setting of the detector axis $\hat{\zdet}$, the momentum distribution $\Pi(E,\theta,\phi)$ will depend on all three variables $E,\theta,\phi$, even though the momentum distribution may often be cylindrically symmetric in some other coordinates. \\

It is desirable to design the detector system so that the detector function can be approximated as the product of an energy acceptance function and an angular acceptance function
\beq
\kappa(\Edet|\myvec{p},\auxpars) \approx \kappa_E(\Edet;E,\auxpars) \Theta(\thmax-\theta) \label{detprod}
\eeq
where $E$, $\Edet$ are the electron energy at the interaction region and the electron energy read out by the detector, $\thmax$ is the detector acceptance angle, and $\Theta(x)$ is the step function defined by $\Theta(x<0) \equiv 0$ and $\Theta(x\geq0) \equiv 1$.  The electron energy spectrum measured by the detector becomes
\beq
Y(\Edet;\auxpars) = \int_0^\mathrm{2 \pi} \!d\phi \int_0^\mathrm{\thmax} \!d\theta \int_0^\infty \!dE\: \sin \theta \: \kappa_E(\Edet;E,\auxpars) \Pi(E,\theta,\phi)
\eeq
In the general case, the dependence of $\Pi$ on $\phi$ prevents further simplification of the angular integral. \\

For an ideal TOF, operating at its default values of the auxiliary parameters $\auxpars_\mathrm{def}$, the energy acceptance function can be approximated by the Dirac delta function
\beq
\kappa_E(\Edet;E,\auxpars_\mathrm{def}) = \delta(\Edet-E)
\eeq
Ideally, the acceptance angle of the TOF, $\thmax$, is almost zero. Then $\Pi(E,\theta,\phi)$ becomes almost independent of $\theta$ and $\phi$. If all these stringent assumptions are satisfied, the detected energy spectrum assumes the simple form
\beq
Y(\Edet;\auxpars_\mathrm{def}) = \pi \thmax^2 \Pi(E,0,0)
\eeq
In reality, the TOF will exhibit some finite energy resolution, which can be determined by electron trajectory simulations. \\

For an ideal repeller detector, there is a single auxiliary parameter $V_D$ of central importance. For a given setting of the deflection voltage $V_D$, the detector measures a single number $Y$. The energy acceptance function can be approximated as
\beq
\kappa_E(\Edet;E,V_D) = \Theta(E - E_c(V_D)) \label{repeldet}
\eeq
where the electron cutoff energy $E_c$ is determined by $V_D$. If we again assume that the acceptance angle $\thmax$ is almost zero, the ideal repeller detector therefore measures
\beq
Y(V_D) = \pi \thmax^2 \int_{E_c(V_D)}^\infty \!dE\: \Pi(E,0,0)
\eeq
and inverting the relationship $E_c = E_c(V_D)$ gives us the yield as a function of cutoff energy, $Y(E_c)$. \\

\begin{figure}[thbp]
\begin{center}
\includegraphics[width=\columnwidth/2]{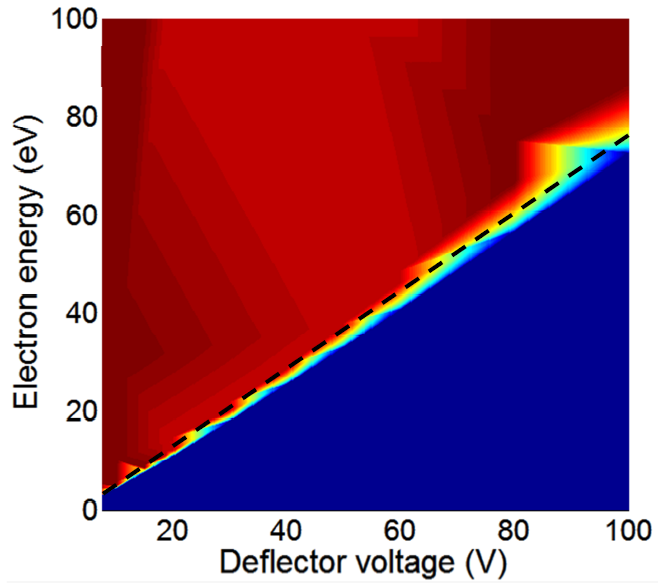}
\caption{Numerical simulation of the acceptance angle for the repeller detector described in Section \ref{sec-repeller}. The colors represent the value of the acceptance angle, from blue (0 mrad) to red (120 mrad). The dotted line marks the approximately linear relationship between cutoff energy and deflection voltage.}
\label{DetectorFunction}
\end{center}
\end{figure}

In reality, the approximation $\thmax \rightarrow 0$ is too stringent: the detection rate falls to zero in this limit. Also, the energy acceptance function is never perfectly sharp. To make accurate predictions of the data in Section \ref{sec-aasfexpts}, it was necessary to estimate the full detection function using simulations of the electron trajectories. These simulations were carried out using the commercial software package \simion. Figure \ref{DetectorFunction} shows the numerically simulated relationship between the deflection voltage, the energy cutoff, and the acceptance angle. The line between blue and red regions marks the approximate relationship $E_c(V_D) = 0.8(V_D - 5)$ discussed in Section \ref{sec-repeller}. The detected yield was calculated using this relationship in conjunction with Eq. (\ref{momdet}). Approximating the acceptance function with a perfectly sharp cutoff led to significant underestimates of the yield at low cutoff energy. \\

\section{Summary and outlook}

Experiments on atomic hydrogen have provided important benchmarks for strong-field ionisation studies, from qualitative interpretations of multiphoton ionisation and the photoelectron plateau, to quantitative theory-experiment comparisons and calibration of laser parameters. Present-day measurements can resolve the differences between exact 3D TDSE simulations and approximate predictions, even for common and reasonable approximations. The achievement of experimental agreement with the 3D TDSE has validated a concrete path to accurate photoelectron measurements with uncertainties at the 10\% level. The methods outlined here are sufficient to find accurate estimates of experimental uncertainty and to convert raw simulation results into accurate predictions of experimental measurements. The laser peak intensity, a key physical parameter for strong-field ionisation, can be accurately retrieved from experimental measurements with 1\% accuracy. \\

Since H is unlikely to be conveniently available in the near future, we envisage that it will serve as a primary reference standard from which secondary calibration standards can be derived. By reaching experiment-theory agreement for H at a given level of uncertainty, one demonstrates the absence of systematic errors at that level. Substituting other atomic and molecular species in the same apparatus, and using otherwise identical experimental techniques, one can obtain new data that is free of systematic errors.  By choosing common reference species like the noble gases or simple molecules, one can also obtain certified reference data with a calibration traceable to H. The reproduction of the reference data in another experiment, in another laboratory, then certifies the accuracy of that experiment without the need to access H directly. Such reference data also poses a stringent test of approximate theories in cases where these theories are the only ones available. \\

We have only begun to explore the potential applications of experiments on H. Measurements like those described here could provide laser intensity calibration at the mid-IR wavelengths characteristic of recent attosecond science experiments. In the mid-IR, the difficulty of standard laser intensity measurements is significantly greater than at 800 nm. Better CEP-resolved measurements could provide a highly accurate calibration of laser CEP by referencing the absolute CEP to 3D-TDSE theory for H. From the point of view of basic physics, the ultimate goal might be a precision measurement of the full electron-ion kinematics of H ionisation using the reaction microscope technique. \\

Improvements in measurement precision generally lead to improved control of systems, and we expect that strong-field ionisation will be no exception. Indeed, for the highly nonlinear physics of strong-field ionisation, the ability to compare control outcomes with exact theory is even greater than usual. In the future, H may well prove to be the ideal testing ground for the control of strong-field processes at the atomic level. \\

\begin{acknowledgments}
The few-cycle measurements described in Section \ref{sec-aasfexpts} represent the achievements of a large experimental and theoretical collaboration. We gratefully acknowledge the outstanding contributions of every team member. \\

The data was collected by the H experimental group at the AASF: Michael Pullen, William Wallace, Omair Ghafur, Dane Laban, Adam Palmer, and Han Xu. The repeller detector system was constructed by Prof Friedrich Hanne of the Westf\"alische Wilhelms-Universit\"at M\"unster. We thank Prof Daniel Kleppner of the Massachusetts Institute of Technology for assistance in setting up the hydrogen source. \\

Theoretical simulations were provided by a large number of groups: Brant Abeln and Prof Klaus Bartschat at Drake University, working with Prof Alexei Grum-Grzhimailo of Lomonosov Moscow State University, Daniel Wells and Prof Harry Quiney of the University of Melbourne, Prof Xiao-Min Tong of the University of Tsukuba, and Igor Ivanov and Prof Anatoli Kheifets of the Australian National University. \\

The present review was supported by the Australian Research Council (ARC) Centre of Excellence for Coherent X-Ray Science (CE0561787) and by the US Air Force Office of Scientific Research under FA2386-12-1-4025. D.K. is supported by an ARC Future Fellowship (FT110100513). \\
\end{acknowledgments}

\appendix

\section{Imperfections of the laser beam}
\label{sec-aberrations}

In reality, the laser beam is never a perfect Gaussian. For most purposes, it appears sufficient to model the beam as a ``simple astigmatic beam'', according to the classification of the International Organization for Standardization (ISO) \cite{ISO-11146-profiling-standard}. In this model, the laser spot is taken to be an elliptical Gaussian with its principal axes along $x$ and $y$. Owing to astigmatism, the foci in the $x$ and $y$ directions may be located at different positions $z_{0,x}$, $z_{0,y}$. Moreover, the divergence of the beam along either $x$ or $y$ may exceed the divergence of an ideal Gaussian beam of the same spot size by the ``beam propagation ratios'' $M_x^2$, $M_y^2$. The intensity becomes (\cite{Yariv-quantum-electronics-BOOK}, ch. 6; \cite{Siegman-beam-quality-tutorial, ISO-11146-profiling-standard})
\ba
I(\myvec{r}) &=& I_0 \frac{w_{0,x}}{w_x(z)} \frac{w_{0,y}}{w_y(z)} \exp \left[ -\frac{x^2}{w_x^2(z)} + \frac{y^2}{w_y^2(z)} \right] \\
w_x(z) &\equiv& w_{0,x} \sqrt{1 + \left( \frac{M_x^2 (z - z_x)}{z_{R,x}} \right)^2}
\ea
Here $w_{0,x}$ is the minimum spot half-width along the $x$ axis at the $x$ focus, with the Rayleigh range $z_{R,x} = \pi w_{0,x}^2/\lambda$, and similarly for $w_{0,y}$. When the beam is well approximated as an elliptical Gaussian ($M_x^2 \approx M_y^2 \approx 1$), the Gouy phase is given by
\beq
\zeta(z) = \frac{1}{2} \left( \tan^{-1} \frac{z - z_x}{z_{R,x}} + \tan^{-1} \frac{z - z_y}{z_{R,y}} \right)
\eeq
However, we are not aware of a simple formula for the Gouy phase for the general case of the simple astigmatic beam. Iterative phase retrieval can be used to compute the Gouy phase from a suitable set of beam intensity distributions acquired at different points along the propagation axis \cite{Laban-Sang-zeptosecond-XUV-interferometer}. \\

Evaluating the beam parameters for the simple astigmatic model is nontrivial. Specifications for commercially available laser systems include information on the beam parameters at the laser output. However, off-axis focusing of the beam by a spherical mirror will introduce further astigmatism. Off-axis paraboloidal mirrors can, in principle, focus without astigmatism, but are highly sensitive to imperfect alignment. \\

\bibliography{C:/Users/Dave/Documents/papers/dkbib}

\begin{thebibliography}{52}%
\makeatletter
\providecommand \@ifxundefined [1]{%
 \@ifx{#1\undefined}
}%
\providecommand \@ifnum [1]{%
 \ifnum #1\expandafter \@firstoftwo
 \else \expandafter \@secondoftwo
 \fi
}%
\providecommand \@ifx [1]{%
 \ifx #1\expandafter \@firstoftwo
 \else \expandafter \@secondoftwo
 \fi
}%
\providecommand \natexlab [1]{#1}%
\providecommand \enquote  [1]{``#1''}%
\providecommand \bibnamefont  [1]{#1}%
\providecommand \bibfnamefont [1]{#1}%
\providecommand \citenamefont [1]{#1}%
\providecommand \href@noop [0]{\@secondoftwo}%
\providecommand \href [0]{\begingroup \@sanitize@url \@href}%
\providecommand \@href[1]{\@@startlink{#1}\@@href}%
\providecommand \@@href[1]{\endgroup#1\@@endlink}%
\providecommand \@sanitize@url [0]{\catcode `\\12\catcode `\$12\catcode
  `\&12\catcode `\#12\catcode `\^12\catcode `\_12\catcode `\%12\relax}%
\providecommand \@@startlink[1]{}%
\providecommand \@@endlink[0]{}%
\providecommand \url  [0]{\begingroup\@sanitize@url \@url }%
\providecommand \@url [1]{\endgroup\@href {#1}{\urlprefix }}%
\providecommand \urlprefix  [0]{URL }%
\providecommand \Eprint [0]{\href }%
\providecommand \doibase [0]{http://dx.doi.org/}%
\providecommand \selectlanguage [0]{\@gobble}%
\providecommand \bibinfo  [0]{\@secondoftwo}%
\providecommand \bibfield  [0]{\@secondoftwo}%
\providecommand \translation [1]{[#1]}%
\providecommand \BibitemOpen [0]{}%
\providecommand \bibitemStop [0]{}%
\providecommand \bibitemNoStop [0]{.\EOS\space}%
\providecommand \EOS [0]{\spacefactor3000\relax}%
\providecommand \BibitemShut  [1]{\csname bibitem#1\endcsname}%
\let\auto@bib@innerbib\@empty
%</preamble>
\bibitem [{\citenamefont {Sheehy}\ and\ \citenamefont
  {DiMauro}(1996)}]{Sheehy-DiMauro-high-field-rev}%
  \BibitemOpen
  \bibfield  {author} {\bibinfo {author} {\bibfnamefont {B}~\bibnamefont
  {Sheehy}}\ and\ \bibinfo {author} {\bibfnamefont {LF}~\bibnamefont
  {DiMauro}},\ }\bibfield  {title} {\enquote {\bibinfo {title} {Atomic and
  molecular dynamics in intense optical fields},}\ }\href@noop {} {\bibfield
  {journal} {\bibinfo  {journal} {Annu. Rev. Phys. Chem.}\ }\textbf {\bibinfo
  {volume} {47}},\ \bibinfo {pages} {463--494} (\bibinfo {year}
  {1996})}\BibitemShut {NoStop}%
\bibitem [{\citenamefont {Corkum}\ and\ \citenamefont
  {Krausz}(2007)}]{Corkum-Krausz-as-rev}%
  \BibitemOpen
  \bibfield  {author} {\bibinfo {author} {\bibfnamefont {P.~B.}\ \bibnamefont
  {Corkum}}\ and\ \bibinfo {author} {\bibfnamefont {F.}~\bibnamefont
  {Krausz}},\ }\bibfield  {title} {\enquote {\bibinfo {title} {Attosecond
  science},}\ }\href@noop {} {\bibfield  {journal} {\bibinfo  {journal} {Nature
  Phys.}\ }\textbf {\bibinfo {volume} {3}},\ \bibinfo {pages} {381--387}
  (\bibinfo {year} {2007})}\BibitemShut {NoStop}%
\bibitem [{\citenamefont {{Joint Committee for Guides in
  Metrology}}(2012)}]{BIPM-international-vocabulary-metrology}%
  \BibitemOpen
  \bibfield  {author} {\bibinfo {author} {\bibnamefont {{Joint Committee for
  Guides in Metrology}}},\ }\href@noop {} {\emph {\bibinfo {title} {JCGM
  200:2012. International vocabulary of metrology –- Basic and general concepts
  and associated terms (VIM)}}},\ \bibinfo {address} {Geneva, Switzerland}
  (\bibinfo {year} {2012})\BibitemShut {NoStop}%
\bibitem [{\citenamefont {{Joint Committee for Guides in
  Metrology}}(2008)}]{BIPM-guide-uncertainty-measurement}%
  \BibitemOpen
  \bibfield  {author} {\bibinfo {author} {\bibnamefont {{Joint Committee for
  Guides in Metrology}}},\ }\href@noop {} {\emph {\bibinfo {title} {JCGM
  100:2008. Evaluation of measurement data -- Guide to the expression of
  uncertainty in measurement}}},\ \bibinfo {address} {Geneva, Switzerland}
  (\bibinfo {year} {2008})\BibitemShut {NoStop}%
\bibitem [{\citenamefont {Rottke}\ \emph {et~al.}(1990)\citenamefont {Rottke},
  \citenamefont {Wolff}, \citenamefont {Brickwedde}, \citenamefont {Feldmann},\
  and\ \citenamefont {Welge}}]{Rottke-Welge-hydrogen-MPI}%
  \BibitemOpen
  \bibfield  {author} {\bibinfo {author} {\bibfnamefont {H}~\bibnamefont
  {Rottke}}, \bibinfo {author} {\bibfnamefont {B}~\bibnamefont {Wolff}},
  \bibinfo {author} {\bibfnamefont {M}~\bibnamefont {Brickwedde}}, \bibinfo
  {author} {\bibfnamefont {D}~\bibnamefont {Feldmann}}, \ and\ \bibinfo
  {author} {\bibfnamefont {KH}~\bibnamefont {Welge}},\ }\bibfield  {title}
  {\enquote {\bibinfo {title} {Multiphoton ionization of atomic hydrogen in
  intense subpicosecond laser pulses},}\ }\href@noop {} {\bibfield  {journal}
  {\bibinfo  {journal} {Phys. Rev. Lett.}\ }\textbf {\bibinfo {volume} {64}},\
  \bibinfo {pages} {404} (\bibinfo {year} {1990})}\BibitemShut {NoStop}%
\bibitem [{\citenamefont {Rottke}\ \emph {et~al.}(1993)\citenamefont {Rottke},
  \citenamefont {Feldmann}, \citenamefont {Wolff-Rottke},\ and\ \citenamefont
  {Welge}}]{Rottke-Welge-hydrogen-low-order-ATI}%
  \BibitemOpen
  \bibfield  {author} {\bibinfo {author} {\bibfnamefont {H}~\bibnamefont
  {Rottke}}, \bibinfo {author} {\bibfnamefont {D}~\bibnamefont {Feldmann}},
  \bibinfo {author} {\bibfnamefont {B}~\bibnamefont {Wolff-Rottke}}, \ and\
  \bibinfo {author} {\bibfnamefont {K~H}\ \bibnamefont {Welge}},\ }\bibfield
  {title} {\enquote {\bibinfo {title} {Atomic hydrogen in intense short laser
  pulses: a new series of photoelectron peaks from above-threshold
  ionization},}\ }\href {http://stacks.iop.org/0953-4075/26/i=2/a=001}
  {\bibfield  {journal} {\bibinfo  {journal} {J. Phys. B}\ }\textbf {\bibinfo
  {volume} {26}},\ \bibinfo {pages} {L15} (\bibinfo {year} {1993})}\BibitemShut
  {NoStop}%
\bibitem [{\citenamefont {Rottke}\ \emph {et~al.}(1994)\citenamefont {Rottke},
  \citenamefont {Wolff-Rottke}, \citenamefont {Feldmann}, \citenamefont
  {Welge}, \citenamefont {D{\"o}rr}, \citenamefont {Potvliege},\ and\
  \citenamefont {Shakeshaft}}]{Rottke-Shakeshaft-hydrogen-low-order-ATI}%
  \BibitemOpen
  \bibfield  {author} {\bibinfo {author} {\bibfnamefont {H}~\bibnamefont
  {Rottke}}, \bibinfo {author} {\bibfnamefont {B}~\bibnamefont {Wolff-Rottke}},
  \bibinfo {author} {\bibfnamefont {D}~\bibnamefont {Feldmann}}, \bibinfo
  {author} {\bibfnamefont {KH}~\bibnamefont {Welge}}, \bibinfo {author}
  {\bibfnamefont {M}~\bibnamefont {D{\"o}rr}}, \bibinfo {author} {\bibfnamefont
  {RM}~\bibnamefont {Potvliege}}, \ and\ \bibinfo {author} {\bibfnamefont
  {R}~\bibnamefont {Shakeshaft}},\ }\bibfield  {title} {\enquote {\bibinfo
  {title} {Atomic hydrogen in a strong optical radiation field},}\ }\href@noop
  {} {\bibfield  {journal} {\bibinfo  {journal} {Phys. Rev. A}\ }\textbf
  {\bibinfo {volume} {49}},\ \bibinfo {pages} {4837} (\bibinfo {year}
  {1994})}\BibitemShut {NoStop}%
\bibitem [{\citenamefont {Paulus}\ \emph {et~al.}(1996)\citenamefont {Paulus},
  \citenamefont {Nicklich}, \citenamefont {Zacher}, \citenamefont
  {Lambropoulos},\ and\ \citenamefont
  {Walther}}]{Paulus-Walther-hydrogen-ATI-expt}%
  \BibitemOpen
  \bibfield  {author} {\bibinfo {author} {\bibfnamefont {G~G}\ \bibnamefont
  {Paulus}}, \bibinfo {author} {\bibfnamefont {W}~\bibnamefont {Nicklich}},
  \bibinfo {author} {\bibfnamefont {F}~\bibnamefont {Zacher}}, \bibinfo
  {author} {\bibfnamefont {P}~\bibnamefont {Lambropoulos}}, \ and\ \bibinfo
  {author} {\bibfnamefont {H}~\bibnamefont {Walther}},\ }\bibfield  {title}
  {\enquote {\bibinfo {title} {High-order above-threshold ionization of atomic
  hydrogen using intense, ultrashort laser pulses},}\ }\href
  {http://stacks.iop.org/0953-4075/29/i=7/a=002} {\bibfield  {journal}
  {\bibinfo  {journal} {J. Phys. B}\ }\textbf {\bibinfo {volume} {29}},\
  \bibinfo {pages} {L249} (\bibinfo {year} {1996})}\BibitemShut {NoStop}%
\bibitem [{\citenamefont {Pullen}\ \emph {et~al.}(2011)\citenamefont {Pullen},
  \citenamefont {Wallace}, \citenamefont {Laban}, \citenamefont {Palmer},
  \citenamefont {Hanne}, \citenamefont {Grum-Grzhimailo}, \citenamefont
  {Abeln}, \citenamefont {Bartschat}, \citenamefont {Weflen}, \citenamefont
  {Ivanov}, \citenamefont {Kheifets}, \citenamefont {Quiney}, \citenamefont
  {Litvinyuk}, \citenamefont {Sang},\ and\ \citenamefont
  {Kielpinski}}]{Pullen-Kielpinski-few-cycle-H-ionisation}%
  \BibitemOpen
  \bibfield  {author} {\bibinfo {author} {\bibfnamefont {M.~G.}\ \bibnamefont
  {Pullen}}, \bibinfo {author} {\bibfnamefont {W.~C.}\ \bibnamefont {Wallace}},
  \bibinfo {author} {\bibfnamefont {D.~E.}\ \bibnamefont {Laban}}, \bibinfo
  {author} {\bibfnamefont {A.~J.}\ \bibnamefont {Palmer}}, \bibinfo {author}
  {\bibfnamefont {G.~F.}\ \bibnamefont {Hanne}}, \bibinfo {author}
  {\bibfnamefont {A.~N.}\ \bibnamefont {Grum-Grzhimailo}}, \bibinfo {author}
  {\bibfnamefont {B.}~\bibnamefont {Abeln}}, \bibinfo {author} {\bibfnamefont
  {K.}~\bibnamefont {Bartschat}}, \bibinfo {author} {\bibfnamefont
  {D.}~\bibnamefont {Weflen}}, \bibinfo {author} {\bibfnamefont
  {I.}~\bibnamefont {Ivanov}}, \bibinfo {author} {\bibfnamefont
  {A.}~\bibnamefont {Kheifets}}, \bibinfo {author} {\bibfnamefont {H.~M.}\
  \bibnamefont {Quiney}}, \bibinfo {author} {\bibfnamefont {I.~V.}\
  \bibnamefont {Litvinyuk}}, \bibinfo {author} {\bibfnamefont {R.~T.}\
  \bibnamefont {Sang}}, \ and\ \bibinfo {author} {\bibfnamefont
  {D.}~\bibnamefont {Kielpinski}},\ }\bibfield  {title} {\enquote {\bibinfo
  {title} {Experimental ionization of atomic hydrogen with few-cycle pulses},}\
  }\href {\doibase 10.1364/OL.36.003660} {\bibfield  {journal} {\bibinfo
  {journal} {Opt. Lett.}\ }\textbf {\bibinfo {volume} {36}},\ \bibinfo {pages}
  {3660--3662} (\bibinfo {year} {2011})}\BibitemShut {NoStop}%
\bibitem [{\citenamefont
  {Pullen}(2011)}]{Pullen-Kielpinski-H-ionisation-THESIS}%
  \BibitemOpen
  \bibfield  {author} {\bibinfo {author} {\bibfnamefont {M.~G.}\ \bibnamefont
  {Pullen}},\ }\emph {\bibinfo {title} {Above threshold ionisation of atomic
  hydrogen using few-cycle pulses}},\ \href@noop {} {Ph.D. thesis},\ \bibinfo
  {school} {Griffith University}, \bibinfo {address} {Brisbane, Australia}
  (\bibinfo {year} {2011}),\ \bibinfo {note} {available at
  http://libraryguides.griffith.edu.au/theses}\BibitemShut {NoStop}%
\bibitem [{\citenamefont {Wallace}\ \emph {et~al.}(2013)\citenamefont
  {Wallace}, \citenamefont {Pullen}, \citenamefont {Laban}, \citenamefont
  {Ghafur}, \citenamefont {Xu}, \citenamefont {Palmer}, \citenamefont {Hanne},
  \citenamefont {Bartschat}, \citenamefont {{Grum-Grzhimailo}}, \citenamefont
  {Quiney}, \citenamefont {Litvinyuk}, \citenamefont {Sang},\ and\
  \citenamefont {Kielpinski}}]{Wallace-Kielpinski-H-ionisation-CEP-effects}%
  \BibitemOpen
  \bibfield  {author} {\bibinfo {author} {\bibfnamefont {W~C}\ \bibnamefont
  {Wallace}}, \bibinfo {author} {\bibfnamefont {M~G}\ \bibnamefont {Pullen}},
  \bibinfo {author} {\bibfnamefont {D~E}\ \bibnamefont {Laban}}, \bibinfo
  {author} {\bibfnamefont {O}~\bibnamefont {Ghafur}}, \bibinfo {author}
  {\bibfnamefont {H}~\bibnamefont {Xu}}, \bibinfo {author} {\bibfnamefont
  {A~J}\ \bibnamefont {Palmer}}, \bibinfo {author} {\bibfnamefont {G~F}\
  \bibnamefont {Hanne}}, \bibinfo {author} {\bibfnamefont {K}~\bibnamefont
  {Bartschat}}, \bibinfo {author} {\bibfnamefont {A~N}\ \bibnamefont
  {{Grum-Grzhimailo}}}, \bibinfo {author} {\bibfnamefont {H~M}\ \bibnamefont
  {Quiney}}, \bibinfo {author} {\bibfnamefont {I~V}\ \bibnamefont {Litvinyuk}},
  \bibinfo {author} {\bibfnamefont {R~T}\ \bibnamefont {Sang}}, \ and\ \bibinfo
  {author} {\bibfnamefont {D}~\bibnamefont {Kielpinski}},\ }\bibfield  {title}
  {\enquote {\bibinfo {title} {Carrier-envelope phase effects in
  above-threshold ionization of atomic hydrogen},}\ }\href@noop {} {\bibfield
  {journal} {\bibinfo  {journal} {New J. Phys.}\ }\textbf {\bibinfo {volume}
  {15}},\ \bibinfo {pages} {033002} (\bibinfo {year} {2013})}\BibitemShut
  {NoStop}%
\bibitem [{\citenamefont {Slevin}\ and\ \citenamefont
  {Stirling}(1981)}]{Slevin-Stirling-rf-H-dissociator}%
  \BibitemOpen
  \bibfield  {author} {\bibinfo {author} {\bibfnamefont {J.}~\bibnamefont
  {Slevin}}\ and\ \bibinfo {author} {\bibfnamefont {W.}~\bibnamefont
  {Stirling}},\ }\bibfield  {title} {\enquote {\bibinfo {title} {Radio
  frequency atomic hydrogen beam source},}\ }\href@noop {} {\bibfield
  {journal} {\bibinfo  {journal} {Rev. Sci. Instrum.}\ }\textbf {\bibinfo
  {volume} {52}},\ \bibinfo {pages} {1780--1782} (\bibinfo {year}
  {1981})}\BibitemShut {NoStop}%
\bibitem [{\citenamefont {Schwonek}(1990)}]{Schwonek-Kleppner-H-source-THESIS}%
  \BibitemOpen
  \bibfield  {author} {\bibinfo {author} {\bibfnamefont {J.~P.}\ \bibnamefont
  {Schwonek}},\ }\emph {\bibinfo {title} {A study of a cold atomic hydrogen
  beam source}},\ \href@noop {} {Ph.D. thesis},\ \bibinfo  {school}
  {Massachusetts Institute of Technology}, \bibinfo {address} {Cambridge, MA,
  USA} (\bibinfo {year} {1990}),\ \bibinfo {note} {available at
  http://hdl.handle.net/1721.1/37496}\BibitemShut {NoStop}%
\bibitem [{\citenamefont {Lavrov}\ \emph
  {et~al.}(2006{\natexlab{a}})\citenamefont {Lavrov}, \citenamefont {Pipa},\
  and\ \citenamefont
  {R{\"o}pcke}}]{Lavrov-Ropcke-emission-spectrum-dissoc-fraction-theory}%
  \BibitemOpen
  \bibfield  {author} {\bibinfo {author} {\bibfnamefont {B.~P.}\ \bibnamefont
  {Lavrov}}, \bibinfo {author} {\bibfnamefont {A.~V.}\ \bibnamefont {Pipa}}, \
  and\ \bibinfo {author} {\bibfnamefont {J.}~\bibnamefont {R{\"o}pcke}},\
  }\bibfield  {title} {\enquote {\bibinfo {title} {On determination of the
  degree of dissociation of hydrogen in non-equilibrium plasmas by means of
  emission spectroscopy: I. the collision-radiative model and numerical
  experiments},}\ }\href@noop {} {\bibfield  {journal} {\bibinfo  {journal}
  {Plasma Sources Sci. Technol.}\ }\textbf {\bibinfo {volume} {15}},\ \bibinfo
  {pages} {135} (\bibinfo {year} {2006}{\natexlab{a}})}\BibitemShut {NoStop}%
\bibitem [{\citenamefont {Lavrov}\ \emph
  {et~al.}(2006{\natexlab{b}})\citenamefont {Lavrov}, \citenamefont {Lang},
  \citenamefont {Pipa},\ and\ \citenamefont
  {R{\"o}pcke}}]{Lavrov-Ropcke-emission-spectrum-dissoc-fraction-expt}%
  \BibitemOpen
  \bibfield  {author} {\bibinfo {author} {\bibfnamefont {B~P}\ \bibnamefont
  {Lavrov}}, \bibinfo {author} {\bibfnamefont {N}~\bibnamefont {Lang}},
  \bibinfo {author} {\bibfnamefont {A~V}\ \bibnamefont {Pipa}}, \ and\ \bibinfo
  {author} {\bibfnamefont {J}~\bibnamefont {R{\"o}pcke}},\ }\bibfield  {title}
  {\enquote {\bibinfo {title} {On determination of the degree of dissociation
  of hydrogen in non-equilibrium plasmas by means of emission spectroscopy:
  \mbox{II. E}xperimental verification},}\ }\href
  {http://stacks.iop.org/0963-0252/15/i=1/a=021} {\bibfield  {journal}
  {\bibinfo  {journal} {Plasma Sources Sci. Technol.}\ }\textbf {\bibinfo
  {volume} {15}},\ \bibinfo {pages} {147} (\bibinfo {year}
  {2006}{\natexlab{b}})}\BibitemShut {NoStop}%
\bibitem [{\citenamefont {Pullen}\ \emph {et~al.}(2013)\citenamefont {Pullen},
  \citenamefont {Wallace}, \citenamefont {Laban}, \citenamefont {Palmer},
  \citenamefont {Hanne}, \citenamefont {{Grum-Grzhimailo}}, \citenamefont
  {Bartschat}, \citenamefont {Ivanov}, \citenamefont {Kheifets}, \citenamefont
  {Wells}, \citenamefont {Quiney}, \citenamefont {Tong}, \citenamefont
  {Litvinyuk}, \citenamefont {Sang},\ and\ \citenamefont
  {Kielpinski}}]{Pullen-Kielpinski-peak-intensity-calibration}%
  \BibitemOpen
  \bibfield  {author} {\bibinfo {author} {\bibfnamefont {MG}~\bibnamefont
  {Pullen}}, \bibinfo {author} {\bibfnamefont {WC}~\bibnamefont {Wallace}},
  \bibinfo {author} {\bibfnamefont {DE}~\bibnamefont {Laban}}, \bibinfo
  {author} {\bibfnamefont {AJ}~\bibnamefont {Palmer}}, \bibinfo {author}
  {\bibfnamefont {GF}~\bibnamefont {Hanne}}, \bibinfo {author} {\bibfnamefont
  {AN}~\bibnamefont {{Grum-Grzhimailo}}}, \bibinfo {author} {\bibfnamefont
  {K}~\bibnamefont {Bartschat}}, \bibinfo {author} {\bibfnamefont
  {I}~\bibnamefont {Ivanov}}, \bibinfo {author} {\bibfnamefont {A}~\bibnamefont
  {Kheifets}}, \bibinfo {author} {\bibfnamefont {D}~\bibnamefont {Wells}},
  \bibinfo {author} {\bibfnamefont {HM}~\bibnamefont {Quiney}}, \bibinfo
  {author} {\bibfnamefont {XM}~\bibnamefont {Tong}}, \bibinfo {author}
  {\bibfnamefont {IV}~\bibnamefont {Litvinyuk}}, \bibinfo {author}
  {\bibfnamefont {RT}~\bibnamefont {Sang}}, \ and\ \bibinfo {author}
  {\bibfnamefont {D}~\bibnamefont {Kielpinski}},\ }\bibfield  {title} {\enquote
  {\bibinfo {title} {Measurement of laser intensities approaching $10^{15}
  \:\mbox{W}/\mbox{cm}^2$ with an accuracy of 1\%},}\ }\href@noop {} {\bibfield
   {journal} {\bibinfo  {journal} {Phys. Rev. A}\ }\textbf {\bibinfo {volume}
  {87}},\ \bibinfo {pages} {053411} (\bibinfo {year} {2013})}\BibitemShut
  {NoStop}%
\bibitem [{\citenamefont {Rathje}\ \emph {et~al.}(2012)\citenamefont {Rathje},
  \citenamefont {Johnson}, \citenamefont {M{\"o}ller}, \citenamefont
  {S{\"u}{\ss}mann}, \citenamefont {Adolph}, \citenamefont {K{\"u}bel},
  \citenamefont {Kienberger}, \citenamefont {Kling}, \citenamefont {Paulus},\
  and\ \citenamefont {Sayler}}]{Rathje-Sayler-ATI-CEP-measurement-rev}%
  \BibitemOpen
  \bibfield  {author} {\bibinfo {author} {\bibfnamefont {T}~\bibnamefont
  {Rathje}}, \bibinfo {author} {\bibfnamefont {NG}~\bibnamefont {Johnson}},
  \bibinfo {author} {\bibfnamefont {M}~\bibnamefont {M{\"o}ller}}, \bibinfo
  {author} {\bibfnamefont {F}~\bibnamefont {S{\"u}{\ss}mann}}, \bibinfo
  {author} {\bibfnamefont {D}~\bibnamefont {Adolph}}, \bibinfo {author}
  {\bibfnamefont {M}~\bibnamefont {K{\"u}bel}}, \bibinfo {author}
  {\bibfnamefont {R}~\bibnamefont {Kienberger}}, \bibinfo {author}
  {\bibfnamefont {MF}~\bibnamefont {Kling}}, \bibinfo {author} {\bibfnamefont
  {GG}~\bibnamefont {Paulus}}, \ and\ \bibinfo {author} {\bibfnamefont
  {AM}~\bibnamefont {Sayler}},\ }\bibfield  {title} {\enquote {\bibinfo {title}
  {Review of attosecond resolved measurement and control via carrier--envelope
  phase tagging with above-threshold ionization},}\ }\href@noop {} {\bibfield
  {journal} {\bibinfo  {journal} {J. Phys. B}\ }\textbf {\bibinfo {volume}
  {45}},\ \bibinfo {pages} {074003} (\bibinfo {year} {2012})}\BibitemShut
  {NoStop}%
\bibitem [{\citenamefont {Paulus}\ \emph {et~al.}(2004)\citenamefont {Paulus},
  \citenamefont {Lindner}, \citenamefont {Milo{\v{s}}evi{\'c}},\ and\
  \citenamefont {Becker}}]{Paulus-Becker-CEP-depdt-ionisation-rev}%
  \BibitemOpen
  \bibfield  {author} {\bibinfo {author} {\bibfnamefont {GG}~\bibnamefont
  {Paulus}}, \bibinfo {author} {\bibfnamefont {F}~\bibnamefont {Lindner}},
  \bibinfo {author} {\bibfnamefont {DB}~\bibnamefont {Milo{\v{s}}evi{\'c}}}, \
  and\ \bibinfo {author} {\bibfnamefont {W}~\bibnamefont {Becker}},\ }\bibfield
   {title} {\enquote {\bibinfo {title} {Phase-controlled single-cycle
  strong-field photoionization},}\ }\href@noop {} {\bibfield  {journal}
  {\bibinfo  {journal} {Phys. Scr.}\ }\textbf {\bibinfo {volume} {2004}},\
  \bibinfo {pages} {120} (\bibinfo {year} {2004})}\BibitemShut {NoStop}%
\bibitem [{\citenamefont {Hansch}\ \emph {et~al.}(1997)\citenamefont {Hansch},
  \citenamefont {Walker},\ and\ \citenamefont
  {Van~Woerkom}}]{Hansch-VanWoerkom-ATI-fast-intensity-dependence}%
  \BibitemOpen
  \bibfield  {author} {\bibinfo {author} {\bibfnamefont {P}~\bibnamefont
  {Hansch}}, \bibinfo {author} {\bibfnamefont {MA}~\bibnamefont {Walker}}, \
  and\ \bibinfo {author} {\bibfnamefont {LD}~\bibnamefont {Van~Woerkom}},\
  }\bibfield  {title} {\enquote {\bibinfo {title} {Resonant hot-electron
  production in above-threshold ionization},}\ }\href@noop {} {\bibfield
  {journal} {\bibinfo  {journal} {Phys. Rev. A}\ }\textbf {\bibinfo {volume}
  {55}},\ \bibinfo {pages} {R2535} (\bibinfo {year} {1997})}\BibitemShut
  {NoStop}%
\bibitem [{\citenamefont {Paulus}\ \emph
  {et~al.}(2001{\natexlab{a}})\citenamefont {Paulus}, \citenamefont {Grasbon},
  \citenamefont {Walther}, \citenamefont {Kopold},\ and\ \citenamefont
  {Becker}}]{Paulus-Becker-channel-closing}%
  \BibitemOpen
  \bibfield  {author} {\bibinfo {author} {\bibfnamefont {GG}~\bibnamefont
  {Paulus}}, \bibinfo {author} {\bibfnamefont {F}~\bibnamefont {Grasbon}},
  \bibinfo {author} {\bibfnamefont {H}~\bibnamefont {Walther}}, \bibinfo
  {author} {\bibfnamefont {R}~\bibnamefont {Kopold}}, \ and\ \bibinfo {author}
  {\bibfnamefont {W}~\bibnamefont {Becker}},\ }\bibfield  {title} {\enquote
  {\bibinfo {title} {Channel-closing-induced resonances in the above-threshold
  ionization plateau},}\ }\href@noop {} {\bibfield  {journal} {\bibinfo
  {journal} {Phys. Rev. A}\ }\textbf {\bibinfo {volume} {64}},\ \bibinfo
  {pages} {021401} (\bibinfo {year} {2001}{\natexlab{a}})}\BibitemShut
  {NoStop}%
\bibitem [{\citenamefont {Larochelle}\ \emph {et~al.}(1998)\citenamefont
  {Larochelle}, \citenamefont {Talebpour},\ and\ \citenamefont
  {Chin}}]{Larochelle-Chin-ADK-ionisation-yield-comparison}%
  \BibitemOpen
  \bibfield  {author} {\bibinfo {author} {\bibfnamefont {SFJ}\ \bibnamefont
  {Larochelle}}, \bibinfo {author} {\bibfnamefont {A}~\bibnamefont
  {Talebpour}}, \ and\ \bibinfo {author} {\bibfnamefont {SL}~\bibnamefont
  {Chin}},\ }\bibfield  {title} {\enquote {\bibinfo {title} {Coulomb effect in
  multiphoton ionization of rare-gas atoms},}\ }\href@noop {} {\bibfield
  {journal} {\bibinfo  {journal} {J. Phys. B}\ }\textbf {\bibinfo {volume}
  {31}},\ \bibinfo {pages} {1215} (\bibinfo {year} {1998})}\BibitemShut
  {NoStop}%
\bibitem [{\citenamefont {Litvinyuk}\ \emph {et~al.}(2003)\citenamefont
  {Litvinyuk}, \citenamefont {Lee}, \citenamefont {Dooley}, \citenamefont
  {Rayner}, \citenamefont {Villeneuve},\ and\ \citenamefont
  {Corkum}}]{Litvinyuk-Corkum-aligned-molecule-ionisation}%
  \BibitemOpen
  \bibfield  {author} {\bibinfo {author} {\bibfnamefont {I~V}\ \bibnamefont
  {Litvinyuk}}, \bibinfo {author} {\bibfnamefont {K~F}\ \bibnamefont {Lee}},
  \bibinfo {author} {\bibfnamefont {P~W}\ \bibnamefont {Dooley}}, \bibinfo
  {author} {\bibfnamefont {D~M}\ \bibnamefont {Rayner}}, \bibinfo {author}
  {\bibfnamefont {D~M}\ \bibnamefont {Villeneuve}}, \ and\ \bibinfo {author}
  {\bibfnamefont {P~B}\ \bibnamefont {Corkum}},\ }\bibfield  {title} {\enquote
  {\bibinfo {title} {Alignment-dependent strong field ionization of
  molecules},}\ }\href@noop {} {\bibfield  {journal} {\bibinfo  {journal}
  {Phys. Rev. Lett.}\ }\textbf {\bibinfo {volume} {90}},\ \bibinfo {pages}
  {233003} (\bibinfo {year} {2003})}\BibitemShut {NoStop}%
\bibitem [{\citenamefont {Alnaser}\ \emph {et~al.}(2004)\citenamefont
  {Alnaser}, \citenamefont {Tong}, \citenamefont {Osipov}, \citenamefont
  {Voss}, \citenamefont {Maharjan}, \citenamefont {Shan}, \citenamefont
  {Chang},\ and\ \citenamefont
  {Cocke}}]{Alnaser-Cocke-peak-intensity-calibration}%
  \BibitemOpen
  \bibfield  {author} {\bibinfo {author} {\bibfnamefont {A.~S.}\ \bibnamefont
  {Alnaser}}, \bibinfo {author} {\bibfnamefont {X.~M.}\ \bibnamefont {Tong}},
  \bibinfo {author} {\bibfnamefont {T.}~\bibnamefont {Osipov}}, \bibinfo
  {author} {\bibfnamefont {S.}~\bibnamefont {Voss}}, \bibinfo {author}
  {\bibfnamefont {C.~M.}\ \bibnamefont {Maharjan}}, \bibinfo {author}
  {\bibfnamefont {B.}~\bibnamefont {Shan}}, \bibinfo {author} {\bibfnamefont
  {Z.}~\bibnamefont {Chang}}, \ and\ \bibinfo {author} {\bibfnamefont {C.~L.}\
  \bibnamefont {Cocke}},\ }\bibfield  {title} {\enquote {\bibinfo {title}
  {Laser-peak-intensity calibration using recoil-ion momentum imaging},}\
  }\href {\doibase 10.1103/PhysRevA.70.023413} {\bibfield  {journal} {\bibinfo
  {journal} {Phys. Rev. A}\ }\textbf {\bibinfo {volume} {70}},\ \bibinfo
  {pages} {023413} (\bibinfo {year} {2004})}\BibitemShut {NoStop}%
\bibitem [{\citenamefont {Smeenk}\ \emph {et~al.}(2011)\citenamefont {Smeenk},
  \citenamefont {Salvail}, \citenamefont {Arissian}, \citenamefont {Corkum},
  \citenamefont {Hebeisen},\ and\ \citenamefont
  {Staudte}}]{Smeenk-Staudte-peak-intensity-calibration}%
  \BibitemOpen
  \bibfield  {author} {\bibinfo {author} {\bibfnamefont {C}~\bibnamefont
  {Smeenk}}, \bibinfo {author} {\bibfnamefont {JZ}~\bibnamefont {Salvail}},
  \bibinfo {author} {\bibfnamefont {L}~\bibnamefont {Arissian}}, \bibinfo
  {author} {\bibfnamefont {PB}~\bibnamefont {Corkum}}, \bibinfo {author}
  {\bibfnamefont {CT}~\bibnamefont {Hebeisen}}, \ and\ \bibinfo {author}
  {\bibfnamefont {A}~\bibnamefont {Staudte}},\ }\bibfield  {title} {\enquote
  {\bibinfo {title} {Precise in-situ measurement of laser pulse intensity using
  strong field ionization},}\ }\href@noop {} {\bibfield  {journal} {\bibinfo
  {journal} {Opt. Express}\ }\textbf {\bibinfo {volume} {19}},\ \bibinfo
  {pages} {9336--9344} (\bibinfo {year} {2011})}\BibitemShut {NoStop}%
\bibitem [{\citenamefont {Micheau}\ \emph {et~al.}(2009)\citenamefont
  {Micheau}, \citenamefont {Chen}, \citenamefont {Le}, \citenamefont
  {Rauschenberger}, \citenamefont {Kling},\ and\ \citenamefont
  {Lin}}]{Micheau-Lin-laser-parameter-photoelectron-retrieval}%
  \BibitemOpen
  \bibfield  {author} {\bibinfo {author} {\bibfnamefont {S}~\bibnamefont
  {Micheau}}, \bibinfo {author} {\bibfnamefont {Z}~\bibnamefont {Chen}},
  \bibinfo {author} {\bibfnamefont {AT}~\bibnamefont {Le}}, \bibinfo {author}
  {\bibfnamefont {J}~\bibnamefont {Rauschenberger}}, \bibinfo {author}
  {\bibfnamefont {MF}~\bibnamefont {Kling}}, \ and\ \bibinfo {author}
  {\bibfnamefont {CD}~\bibnamefont {Lin}},\ }\bibfield  {title} {\enquote
  {\bibinfo {title} {Accurate retrieval of target structures and laser
  parameters of few-cycle pulses from photoelectron momentum spectra},}\
  }\href@noop {} {\bibfield  {journal} {\bibinfo  {journal} {Phys. Rev. Lett.}\
  }\textbf {\bibinfo {volume} {102}},\ \bibinfo {pages} {073001} (\bibinfo
  {year} {2009})}\BibitemShut {NoStop}%
\bibitem [{\citenamefont {Chen}\ \emph {et~al.}(2009)\citenamefont {Chen},
  \citenamefont {Wittmann}, \citenamefont {Horvath},\ and\ \citenamefont
  {Lin}}]{Chen-Lin-CEP-ATI-waveform-retrieval}%
  \BibitemOpen
  \bibfield  {author} {\bibinfo {author} {\bibfnamefont {Z}~\bibnamefont
  {Chen}}, \bibinfo {author} {\bibfnamefont {T}~\bibnamefont {Wittmann}},
  \bibinfo {author} {\bibfnamefont {B}~\bibnamefont {Horvath}}, \ and\ \bibinfo
  {author} {\bibfnamefont {CD}~\bibnamefont {Lin}},\ }\bibfield  {title}
  {\enquote {\bibinfo {title} {Complete real-time temporal waveform
  characterization of single-shot few-cycle laser pulses},}\ }\href@noop {}
  {\bibfield  {journal} {\bibinfo  {journal} {Phys. Rev. A}\ }\textbf {\bibinfo
  {volume} {80}},\ \bibinfo {pages} {061402} (\bibinfo {year}
  {2009})}\BibitemShut {NoStop}%
\bibitem [{\citenamefont {Taylor}(1996)}]{Taylor-error-analysis-BOOK}%
  \BibitemOpen
  \bibfield  {author} {\bibinfo {author} {\bibfnamefont {J.R.}\ \bibnamefont
  {Taylor}},\ }\href@noop {} {\emph {\bibinfo {title} {An Introduction to Error
  Analysis}}},\ \bibinfo {edition} {2nd}\ ed.\ (\bibinfo  {publisher}
  {University Science Books},\ \bibinfo {address} {Sausalito, CA, USA},\
  \bibinfo {year} {1996})\BibitemShut {NoStop}%
\bibitem [{\citenamefont {Allan}(1966)}]{Allan-deviation-CLASSIC}%
  \BibitemOpen
  \bibfield  {author} {\bibinfo {author} {\bibfnamefont {D~W}\ \bibnamefont
  {Allan}},\ }\bibfield  {title} {\enquote {\bibinfo {title} {Statistics of
  atomic frequency standards},}\ }\href@noop {} {\bibfield  {journal} {\bibinfo
   {journal} {Proc. IEEE}\ }\textbf {\bibinfo {volume} {54}},\ \bibinfo {pages}
  {221--230} (\bibinfo {year} {1966})}\BibitemShut {NoStop}%
\bibitem [{\citenamefont {Barnes}\ \emph {et~al.}(1971)\citenamefont {Barnes},
  \citenamefont {Chi}, \citenamefont {Cutler}, \citenamefont {Healey},
  \citenamefont {Leeson}, \citenamefont {McGunigal}, \citenamefont {Mullen},
  \citenamefont {Smith}, \citenamefont {Sydnor}, \citenamefont {Vessot},\ and\
  \citenamefont {Winkler}}]{Barnes-frequency-stability-allan-variance-REV}%
  \BibitemOpen
  \bibfield  {author} {\bibinfo {author} {\bibfnamefont {J~A}\ \bibnamefont
  {Barnes}}, \bibinfo {author} {\bibfnamefont {A~R}\ \bibnamefont {Chi}},
  \bibinfo {author} {\bibfnamefont {L~S}\ \bibnamefont {Cutler}}, \bibinfo
  {author} {\bibfnamefont {D~J}\ \bibnamefont {Healey}}, \bibinfo {author}
  {\bibfnamefont {D~B}\ \bibnamefont {Leeson}}, \bibinfo {author}
  {\bibfnamefont {T~E}\ \bibnamefont {McGunigal}}, \bibinfo {author}
  {\bibfnamefont {J~A}\ \bibnamefont {Mullen}}, \bibinfo {author}
  {\bibfnamefont {W~L}\ \bibnamefont {Smith}}, \bibinfo {author} {\bibfnamefont
  {R~L}\ \bibnamefont {Sydnor}}, \bibinfo {author} {\bibfnamefont {R~F~C}\
  \bibnamefont {Vessot}}, \ and\ \bibinfo {author} {\bibfnamefont {G~M~R}\
  \bibnamefont {Winkler}},\ }\bibfield  {title} {\enquote {\bibinfo {title}
  {Characterization of frequency stability},}\ }\href@noop {} {\bibfield
  {journal} {\bibinfo  {journal} {IEEE Trans. Instrum. Meas.}\ }\textbf
  {\bibinfo {volume} {1001}},\ \bibinfo {pages} {105--120} (\bibinfo {year}
  {1971})}\BibitemShut {NoStop}%
\bibitem [{\citenamefont {Allan}(1987)}]{Allan-variance-general}%
  \BibitemOpen
  \bibfield  {author} {\bibinfo {author} {\bibfnamefont {D~W}\ \bibnamefont
  {Allan}},\ }\bibfield  {title} {\enquote {\bibinfo {title} {Should the
  classical variance be used as a basic measure in standards metrology?}}\
  }\href@noop {} {\bibfield  {journal} {\bibinfo  {journal} {IEEE Trans.
  Instrum. Meas.}\ }\textbf {\bibinfo {volume} {1001}},\ \bibinfo {pages}
  {646--654} (\bibinfo {year} {1987})}\BibitemShut {NoStop}%
\bibitem [{\citenamefont {Witt}(2009)}]{Witt-allan-variance-tutorial}%
  \BibitemOpen
  \bibfield  {author} {\bibinfo {author} {\bibfnamefont {T.~J.}\ \bibnamefont
  {Witt}},\ }\bibfield  {title} {\enquote {\bibinfo {title} {Practical methods
  for treating serial correlations in experimental observations},}\ }\href
  {\doibase 10.1140/epjst/e2009-01047-1} {\bibfield  {journal} {\bibinfo
  {journal} {Eur. Phys. J.: Spec. Top.}\ }\textbf {\bibinfo {volume} {172}},\
  \bibinfo {pages} {137--152} (\bibinfo {year} {2009})}\BibitemShut {NoStop}%
\bibitem [{\citenamefont {Hopcroft}(2010)}]{Hopcroft-allan-variance-matlab}%
  \BibitemOpen
  \bibfield  {author} {\bibinfo {author} {\bibfnamefont {M.~A.}\ \bibnamefont
  {Hopcroft}},\ }\href@noop {} {}\bibinfo {howpublished} {available at
  http://www.mathworks.com.au/matlabcentral/fileexchange/13246-allan} (\bibinfo
  {year} {2010})\BibitemShut {NoStop}%
\bibitem [{\citenamefont {Laban}\ \emph {et~al.}(2012)\citenamefont {Laban},
  \citenamefont {Palmer}, \citenamefont {Wallace}, \citenamefont {Gaffney},
  \citenamefont {Notermans}, \citenamefont {Clevis}, \citenamefont {Pullen},
  \citenamefont {Jiang}, \citenamefont {Quiney}, \citenamefont {Litvinyuk},
  \citenamefont {Kielpinski},\ and\ \citenamefont
  {Sang}}]{Laban-Sang-zeptosecond-XUV-interferometer}%
  \BibitemOpen
  \bibfield  {author} {\bibinfo {author} {\bibfnamefont {DE}~\bibnamefont
  {Laban}}, \bibinfo {author} {\bibfnamefont {AJ}~\bibnamefont {Palmer}},
  \bibinfo {author} {\bibfnamefont {WC}~\bibnamefont {Wallace}}, \bibinfo
  {author} {\bibfnamefont {NS}~\bibnamefont {Gaffney}}, \bibinfo {author}
  {\bibfnamefont {RPMJW}\ \bibnamefont {Notermans}}, \bibinfo {author}
  {\bibfnamefont {TTJ}\ \bibnamefont {Clevis}}, \bibinfo {author}
  {\bibfnamefont {MG}~\bibnamefont {Pullen}}, \bibinfo {author} {\bibfnamefont
  {D}~\bibnamefont {Jiang}}, \bibinfo {author} {\bibfnamefont {HM}~\bibnamefont
  {Quiney}}, \bibinfo {author} {\bibfnamefont {IV}~\bibnamefont {Litvinyuk}},
  \bibinfo {author} {\bibfnamefont {D}~\bibnamefont {Kielpinski}}, \ and\
  \bibinfo {author} {\bibfnamefont {RT}~\bibnamefont {Sang}},\ }\bibfield
  {title} {\enquote {\bibinfo {title} {Extreme ultraviolet interferometer using
  high-order harmonic generation from successive sources},}\ }\href@noop {}
  {\bibfield  {journal} {\bibinfo  {journal} {Phys. Rev. Lett.}\ }\textbf
  {\bibinfo {volume} {109}},\ \bibinfo {pages} {263902} (\bibinfo {year}
  {2012})}\BibitemShut {NoStop}%
\bibitem [{\citenamefont {Paschotta}(2008)}]{Paschotta-laser-encyclopedia}%
  \BibitemOpen
  \bibfield  {author} {\bibinfo {author} {\bibfnamefont {R.}~\bibnamefont
  {Paschotta}},\ }\href@noop {} {\emph {\bibinfo {title} {Encyclopedia of Laser
  Physics and Technology}}}\ (\bibinfo  {publisher} {Wiley-VCH},\ \bibinfo
  {address} {Berlin},\ \bibinfo {year} {2008})\BibitemShut {NoStop}%
\bibitem [{\citenamefont {Shanmugan}\ and\ \citenamefont
  {Breipohl}(1988)}]{Shanmugan-Breipohl-random-signal-analysis-BOOK}%
  \BibitemOpen
  \bibfield  {author} {\bibinfo {author} {\bibfnamefont {K~S}\ \bibnamefont
  {Shanmugan}}\ and\ \bibinfo {author} {\bibfnamefont {A~M}\ \bibnamefont
  {Breipohl}},\ }\href@noop {} {\emph {\bibinfo {title} {Random Signals:
  Detection, Estimation, and Data Analysis}}}\ (\bibinfo  {publisher} {Wiley},\
  \bibinfo {year} {1988})\BibitemShut {NoStop}%
\bibitem [{\citenamefont {Kahra}\ \emph {et~al.}(2012)\citenamefont {Kahra},
  \citenamefont {Leschhorn}, \citenamefont {Kowalewski}, \citenamefont
  {Schiffrin}, \citenamefont {Bothschafter}, \citenamefont {Fu{\ss}},
  \citenamefont {de~Vivie-Riedle}, \citenamefont {Ernstorfer}, \citenamefont
  {Krausz}, \citenamefont {Kienberger},\ and\ \citenamefont
  {Schaetz}}]{Kahra-Schaetz-single-ion-strong-field}%
  \BibitemOpen
  \bibfield  {author} {\bibinfo {author} {\bibfnamefont {S}~\bibnamefont
  {Kahra}}, \bibinfo {author} {\bibfnamefont {G}~\bibnamefont {Leschhorn}},
  \bibinfo {author} {\bibfnamefont {M}~\bibnamefont {Kowalewski}}, \bibinfo
  {author} {\bibfnamefont {A}~\bibnamefont {Schiffrin}}, \bibinfo {author}
  {\bibfnamefont {E}~\bibnamefont {Bothschafter}}, \bibinfo {author}
  {\bibfnamefont {W}~\bibnamefont {Fu{\ss}}}, \bibinfo {author} {\bibfnamefont
  {R}~\bibnamefont {de~Vivie-Riedle}}, \bibinfo {author} {\bibfnamefont
  {R}~\bibnamefont {Ernstorfer}}, \bibinfo {author} {\bibfnamefont
  {F}~\bibnamefont {Krausz}}, \bibinfo {author} {\bibfnamefont {R}~\bibnamefont
  {Kienberger}}, \ and\ \bibinfo {author} {\bibfnamefont {T}~\bibnamefont
  {Schaetz}},\ }\bibfield  {title} {\enquote {\bibinfo {title} {A molecular
  conveyor belt by controlled delivery of single molecules into ultrashort
  laser pulses},}\ }\href@noop {} {\bibfield  {journal} {\bibinfo  {journal}
  {Nature Phys.}\ }\textbf {\bibinfo {volume} {8}},\ \bibinfo {pages}
  {238--242} (\bibinfo {year} {2012})}\BibitemShut {NoStop}%
\bibitem [{\citenamefont {Saleh}\ and\ \citenamefont
  {Teich}(2007)}]{Saleh-Teich-fundamentals-photonics-BOOK}%
  \BibitemOpen
  \bibfield  {author} {\bibinfo {author} {\bibfnamefont {B.~E.~A.}\
  \bibnamefont {Saleh}}\ and\ \bibinfo {author} {\bibfnamefont {M.~C.}\
  \bibnamefont {Teich}},\ }\href@noop {} {\emph {\bibinfo {title} {Fundamentals
  of Photonics}}},\ \bibinfo {edition} {2nd}\ ed.\ (\bibinfo  {publisher}
  {Wiley},\ \bibinfo {address} {Hoboken, NJ, USA},\ \bibinfo {year}
  {2007})\BibitemShut {NoStop}%
\bibitem [{\citenamefont {Lindner}\ \emph {et~al.}(2004)\citenamefont
  {Lindner}, \citenamefont {Paulus}, \citenamefont {Walther}, \citenamefont
  {Baltu{\v{s}}ka}, \citenamefont {Goulielmakis}, \citenamefont {Lezius},\ and\
  \citenamefont {Krausz}}]{Lindner-Krausz-few-cycle-Gouy-phase}%
  \BibitemOpen
  \bibfield  {author} {\bibinfo {author} {\bibfnamefont {F}~\bibnamefont
  {Lindner}}, \bibinfo {author} {\bibfnamefont {GG}~\bibnamefont {Paulus}},
  \bibinfo {author} {\bibfnamefont {H}~\bibnamefont {Walther}}, \bibinfo
  {author} {\bibfnamefont {A}~\bibnamefont {Baltu{\v{s}}ka}}, \bibinfo {author}
  {\bibfnamefont {E}~\bibnamefont {Goulielmakis}}, \bibinfo {author}
  {\bibfnamefont {M}~\bibnamefont {Lezius}}, \ and\ \bibinfo {author}
  {\bibfnamefont {F}~\bibnamefont {Krausz}},\ }\bibfield  {title} {\enquote
  {\bibinfo {title} {Gouy phase shift for few-cycle laser pulses},}\
  }\href@noop {} {\bibfield  {journal} {\bibinfo  {journal} {Phys. Rev. Lett.}\
  }\textbf {\bibinfo {volume} {92}},\ \bibinfo {pages} {113001--113001}
  (\bibinfo {year} {2004})}\BibitemShut {NoStop}%
\bibitem [{\citenamefont {Boutu}\ \emph {et~al.}(2011)\citenamefont {Boutu},
  \citenamefont {Auguste}, \citenamefont {Boyko}, \citenamefont {Sola},
  \citenamefont {Balcou}, \citenamefont {Binazon}, \citenamefont {Gobert},
  \citenamefont {Merdji}, \citenamefont {Valentin}, \citenamefont {Constant},
  \citenamefont {M{\'e}vel},\ and\ \citenamefont
  {Carr{\'e}}}]{Boutu-Carre-flat-top-beam}%
  \BibitemOpen
  \bibfield  {author} {\bibinfo {author} {\bibfnamefont {W}~\bibnamefont
  {Boutu}}, \bibinfo {author} {\bibfnamefont {T}~\bibnamefont {Auguste}},
  \bibinfo {author} {\bibfnamefont {O}~\bibnamefont {Boyko}}, \bibinfo {author}
  {\bibfnamefont {I}~\bibnamefont {Sola}}, \bibinfo {author} {\bibfnamefont
  {P}~\bibnamefont {Balcou}}, \bibinfo {author} {\bibfnamefont {L}~\bibnamefont
  {Binazon}}, \bibinfo {author} {\bibfnamefont {O}~\bibnamefont {Gobert}},
  \bibinfo {author} {\bibfnamefont {H}~\bibnamefont {Merdji}}, \bibinfo
  {author} {\bibfnamefont {C}~\bibnamefont {Valentin}}, \bibinfo {author}
  {\bibfnamefont {E}~\bibnamefont {Constant}}, \bibinfo {author} {\bibfnamefont
  {E}~\bibnamefont {M{\'e}vel}}, \ and\ \bibinfo {author} {\bibfnamefont
  {B}~\bibnamefont {Carr{\'e}}},\ }\bibfield  {title} {\enquote {\bibinfo
  {title} {High-order-harmonic generation in gas with a flat-top laser beam},}\
  }\href@noop {} {\bibfield  {journal} {\bibinfo  {journal} {Phys. Rev. A}\
  }\textbf {\bibinfo {volume} {84}},\ \bibinfo {pages} {063406} (\bibinfo
  {year} {2011})}\BibitemShut {NoStop}%
\bibitem [{\citenamefont {Fu}\ \emph {et~al.}(2009)\citenamefont {Fu},
  \citenamefont {Xiong}, \citenamefont {Xu}, \citenamefont {Yao}, \citenamefont
  {Zeng}, \citenamefont {Chu}, \citenamefont {Cheng}, \citenamefont {Xu},
  \citenamefont {Liu},\ and\ \citenamefont {Chin}}]{Fu-Chin-flat-top-beam}%
  \BibitemOpen
  \bibfield  {author} {\bibinfo {author} {\bibfnamefont {Y}~\bibnamefont {Fu}},
  \bibinfo {author} {\bibfnamefont {H}~\bibnamefont {Xiong}}, \bibinfo {author}
  {\bibfnamefont {H}~\bibnamefont {Xu}}, \bibinfo {author} {\bibfnamefont
  {J}~\bibnamefont {Yao}}, \bibinfo {author} {\bibfnamefont {B}~\bibnamefont
  {Zeng}}, \bibinfo {author} {\bibfnamefont {W}~\bibnamefont {Chu}}, \bibinfo
  {author} {\bibfnamefont {Y}~\bibnamefont {Cheng}}, \bibinfo {author}
  {\bibfnamefont {Z}~\bibnamefont {Xu}}, \bibinfo {author} {\bibfnamefont
  {W}~\bibnamefont {Liu}}, \ and\ \bibinfo {author} {\bibfnamefont {S~L}\
  \bibnamefont {Chin}},\ }\bibfield  {title} {\enquote {\bibinfo {title}
  {Generation of extended filaments of femtosecond pulses in air by use of a
  single-step phase plate},}\ }\href@noop {} {\bibfield  {journal} {\bibinfo
  {journal} {Opt. Lett.}\ }\textbf {\bibinfo {volume} {34}},\ \bibinfo {pages}
  {3752--3754} (\bibinfo {year} {2009})}\BibitemShut {NoStop}%
\bibitem [{\citenamefont {Baltu{\v{s}}ka}\ \emph {et~al.}(2003)\citenamefont
  {Baltu{\v{s}}ka}, \citenamefont {Uiberacker}, \citenamefont {Goulielmakis},
  \citenamefont {Kienberger}, \citenamefont {Yakovlev}, \citenamefont {Udem},
  \citenamefont {H{\"a}nsch},\ and\ \citenamefont
  {Krausz}}]{Baltuska-Krausz-phase-stable-amplifier}%
  \BibitemOpen
  \bibfield  {author} {\bibinfo {author} {\bibfnamefont {A}~\bibnamefont
  {Baltu{\v{s}}ka}}, \bibinfo {author} {\bibfnamefont {M}~\bibnamefont
  {Uiberacker}}, \bibinfo {author} {\bibfnamefont {E}~\bibnamefont
  {Goulielmakis}}, \bibinfo {author} {\bibfnamefont {R}~\bibnamefont
  {Kienberger}}, \bibinfo {author} {\bibfnamefont {V~S}\ \bibnamefont
  {Yakovlev}}, \bibinfo {author} {\bibfnamefont {T}~\bibnamefont {Udem}},
  \bibinfo {author} {\bibfnamefont {T~W}\ \bibnamefont {H{\"a}nsch}}, \ and\
  \bibinfo {author} {\bibfnamefont {F}~\bibnamefont {Krausz}},\ }\bibfield
  {title} {\enquote {\bibinfo {title} {Phase-controlled amplification of
  few-cycle laser pulses},}\ }\href@noop {} {\bibfield  {journal} {\bibinfo
  {journal} {IEEE J. Sel. Top. Quant. Elec.}\ }\textbf {\bibinfo {volume}
  {9}},\ \bibinfo {pages} {972--989} (\bibinfo {year} {2003})}\BibitemShut
  {NoStop}%
\bibitem [{\citenamefont {Adolph}\ \emph {et~al.}(2011)\citenamefont {Adolph},
  \citenamefont {Sayler}, \citenamefont {Rathje}, \citenamefont {R\"{u}hle},\
  and\ \citenamefont {Paulus}}]{Adolph-Paulus-stereo-ATI-CEP-locking}%
  \BibitemOpen
  \bibfield  {author} {\bibinfo {author} {\bibfnamefont {D.}~\bibnamefont
  {Adolph}}, \bibinfo {author} {\bibfnamefont {A.~M.}\ \bibnamefont {Sayler}},
  \bibinfo {author} {\bibfnamefont {T.}~\bibnamefont {Rathje}}, \bibinfo
  {author} {\bibfnamefont {K.}~\bibnamefont {R\"{u}hle}}, \ and\ \bibinfo
  {author} {\bibfnamefont {G.~G.}\ \bibnamefont {Paulus}},\ }\bibfield  {title}
  {\enquote {\bibinfo {title} {Improved carrier-envelope phase locking of
  intense few-cycle laser pulses using above-threshold ionization},}\ }\href
  {\doibase 10.1364/OL.36.003639} {\bibfield  {journal} {\bibinfo  {journal}
  {Opt. Lett.}\ }\textbf {\bibinfo {volume} {36}},\ \bibinfo {pages}
  {3639--3641} (\bibinfo {year} {2011})}\BibitemShut {NoStop}%
\bibitem [{\citenamefont {Vozzi}\ \emph {et~al.}(2006)\citenamefont {Vozzi},
  \citenamefont {Cirmi}, \citenamefont {Manzoni}, \citenamefont {Benedetti},
  \citenamefont {Calegari}, \citenamefont {Sansone}, \citenamefont {Stagira},
  \citenamefont {Svelto}, \citenamefont {De~Silvestri}, \citenamefont
  {Nisoli},\ and\ \citenamefont
  {Cerullo}}]{Vozzi-Cerullo-CEP-stable-midIR-amplifier}%
  \BibitemOpen
  \bibfield  {author} {\bibinfo {author} {\bibfnamefont {C}~\bibnamefont
  {Vozzi}}, \bibinfo {author} {\bibfnamefont {G}~\bibnamefont {Cirmi}},
  \bibinfo {author} {\bibfnamefont {C}~\bibnamefont {Manzoni}}, \bibinfo
  {author} {\bibfnamefont {E}~\bibnamefont {Benedetti}}, \bibinfo {author}
  {\bibfnamefont {F}~\bibnamefont {Calegari}}, \bibinfo {author} {\bibfnamefont
  {G}~\bibnamefont {Sansone}}, \bibinfo {author} {\bibfnamefont
  {S}~\bibnamefont {Stagira}}, \bibinfo {author} {\bibfnamefont
  {O}~\bibnamefont {Svelto}}, \bibinfo {author} {\bibfnamefont {S}~\bibnamefont
  {De~Silvestri}}, \bibinfo {author} {\bibfnamefont {M}~\bibnamefont {Nisoli}},
  \ and\ \bibinfo {author} {\bibfnamefont {G}~\bibnamefont {Cerullo}},\
  }\bibfield  {title} {\enquote {\bibinfo {title} {High-energy,
  few-optical-cycle pulses at 1.5 {$\mu$}m with passive carrier-envelope phase
  stabilization},}\ }\href@noop {} {\bibfield  {journal} {\bibinfo  {journal}
  {Opt. Express}\ }\textbf {\bibinfo {volume} {14}},\ \bibinfo {pages}
  {10109--10116} (\bibinfo {year} {2006})}\BibitemShut {NoStop}%
\bibitem [{\citenamefont {Wittmann}\ \emph {et~al.}(2009)\citenamefont
  {Wittmann}, \citenamefont {Horvath}, \citenamefont {Helml}, \citenamefont
  {Sch{\"a}tzel}, \citenamefont {Gu}, \citenamefont {Cavalieri}, \citenamefont
  {Paulus},\ and\ \citenamefont
  {Kienberger}}]{Wittmann-Kienberger-single-shot-CEO-meast}%
  \BibitemOpen
  \bibfield  {author} {\bibinfo {author} {\bibfnamefont {T.}~\bibnamefont
  {Wittmann}}, \bibinfo {author} {\bibfnamefont {B.}~\bibnamefont {Horvath}},
  \bibinfo {author} {\bibfnamefont {W.}~\bibnamefont {Helml}}, \bibinfo
  {author} {\bibfnamefont {M.~G.}\ \bibnamefont {Sch{\"a}tzel}}, \bibinfo
  {author} {\bibfnamefont {X.}~\bibnamefont {Gu}}, \bibinfo {author}
  {\bibfnamefont {A.~L.}\ \bibnamefont {Cavalieri}}, \bibinfo {author}
  {\bibfnamefont {G.~G.}\ \bibnamefont {Paulus}}, \ and\ \bibinfo {author}
  {\bibfnamefont {R.}~\bibnamefont {Kienberger}},\ }\bibfield  {title}
  {\enquote {\bibinfo {title} {Single-shot carrier\mbox{-}envelope phase
  measurement of few-cycle laser pulses},}\ }\href@noop {} {\bibfield
  {journal} {\bibinfo  {journal} {Nature Phys.}\ }\textbf {\bibinfo {volume}
  {5}},\ \bibinfo {pages} {357--362} (\bibinfo {year} {2009})}\BibitemShut
  {NoStop}%
\bibitem [{\citenamefont {Roudnev}\ and\ \citenamefont
  {Esry}(2007)}]{Roudnev-Esry-general-CEP-theory}%
  \BibitemOpen
  \bibfield  {author} {\bibinfo {author} {\bibfnamefont {V.}~\bibnamefont
  {Roudnev}}\ and\ \bibinfo {author} {\bibfnamefont {B.~D.}\ \bibnamefont
  {Esry}},\ }\bibfield  {title} {\enquote {\bibinfo {title} {General theory of
  carrier-envelope phase effects},}\ }\href {\doibase
  10.1103/PhysRevLett.99.220406} {\bibfield  {journal} {\bibinfo  {journal}
  {Phys. Rev. Lett.}\ }\textbf {\bibinfo {volume} {99}},\ \bibinfo {pages}
  {220406} (\bibinfo {year} {2007})}\BibitemShut {NoStop}%
\bibitem [{\citenamefont {Paulus}\ \emph
  {et~al.}(2001{\natexlab{b}})\citenamefont {Paulus}, \citenamefont {Grasbon},
  \citenamefont {Walther}, \citenamefont {Villoresi}, \citenamefont {Nisoli},
  \citenamefont {Stagira}, \citenamefont {Priori},\ and\ \citenamefont
  {De~Silvestri}}]{Paulus-DeSilvestri-CEO-depdt-ionization}%
  \BibitemOpen
  \bibfield  {author} {\bibinfo {author} {\bibfnamefont {GG}~\bibnamefont
  {Paulus}}, \bibinfo {author} {\bibfnamefont {F.}~\bibnamefont {Grasbon}},
  \bibinfo {author} {\bibfnamefont {H.}~\bibnamefont {Walther}}, \bibinfo
  {author} {\bibfnamefont {P.}~\bibnamefont {Villoresi}}, \bibinfo {author}
  {\bibfnamefont {M.}~\bibnamefont {Nisoli}}, \bibinfo {author} {\bibfnamefont
  {S.}~\bibnamefont {Stagira}}, \bibinfo {author} {\bibfnamefont
  {E.}~\bibnamefont {Priori}}, \ and\ \bibinfo {author} {\bibfnamefont
  {S.}~\bibnamefont {De~Silvestri}},\ }\bibfield  {title} {\enquote {\bibinfo
  {title} {Absolute-phase phenomena in photoionization with few-cycle laser
  pulses},}\ }\href@noop {} {\bibfield  {journal} {\bibinfo  {journal}
  {Nature}\ }\textbf {\bibinfo {volume} {414}},\ \bibinfo {pages} {182--184}
  (\bibinfo {year} {2001}{\natexlab{b}})}\BibitemShut {NoStop}%
\bibitem [{\citenamefont {Paulus}\ \emph {et~al.}(2003)\citenamefont {Paulus},
  \citenamefont {Lindner}, \citenamefont {Walther}, \citenamefont
  {Baltu\ifmmode~\check{s}\else \v{s}\fi{}ka}, \citenamefont {Goulielmakis},
  \citenamefont {Lezius},\ and\ \citenamefont
  {Krausz}}]{Paulus-Krausz-phase-depdt-electron-emission}%
  \BibitemOpen
  \bibfield  {author} {\bibinfo {author} {\bibfnamefont {G.~G.}\ \bibnamefont
  {Paulus}}, \bibinfo {author} {\bibfnamefont {F.}~\bibnamefont {Lindner}},
  \bibinfo {author} {\bibfnamefont {H.}~\bibnamefont {Walther}}, \bibinfo
  {author} {\bibfnamefont {A.}~\bibnamefont {Baltu\ifmmode~\check{s}\else
  \v{s}\fi{}ka}}, \bibinfo {author} {\bibfnamefont {E.}~\bibnamefont
  {Goulielmakis}}, \bibinfo {author} {\bibfnamefont {M.}~\bibnamefont
  {Lezius}}, \ and\ \bibinfo {author} {\bibfnamefont {F.}~\bibnamefont
  {Krausz}},\ }\bibfield  {title} {\enquote {\bibinfo {title} {Measurement of
  the phase of few-cycle laser pulses},}\ }\href {\doibase
  10.1103/PhysRevLett.91.253004} {\bibfield  {journal} {\bibinfo  {journal}
  {Phys. Rev. Lett.}\ }\textbf {\bibinfo {volume} {91}},\ \bibinfo {pages}
  {253004} (\bibinfo {year} {2003})}\BibitemShut {NoStop}%
\bibitem [{\citenamefont {Augst}\ \emph {et~al.}(1991)\citenamefont {Augst},
  \citenamefont {Meyerhofer}, \citenamefont {Strickland},\ and\ \citenamefont
  {Chin}}]{Augst-Chin-barrier-suppression-ionisation}%
  \BibitemOpen
  \bibfield  {author} {\bibinfo {author} {\bibfnamefont {S}~\bibnamefont
  {Augst}}, \bibinfo {author} {\bibfnamefont {D~D}\ \bibnamefont {Meyerhofer}},
  \bibinfo {author} {\bibfnamefont {D}~\bibnamefont {Strickland}}, \ and\
  \bibinfo {author} {\bibfnamefont {SL}~\bibnamefont {Chin}},\ }\bibfield
  {title} {\enquote {\bibinfo {title} {Laser ionization of noble gases by
  coulomb-barrier suppression},}\ }\href@noop {} {\bibfield  {journal}
  {\bibinfo  {journal} {JOSA B}\ }\textbf {\bibinfo {volume} {8}},\ \bibinfo
  {pages} {858--867} (\bibinfo {year} {1991})}\BibitemShut {NoStop}%
\bibitem [{\citenamefont {Hansch}\ and\ \citenamefont
  {Van~Woerkom}(1996)}]{Hansch-VanWoerkom-slit-selection-interaction-region}%
  \BibitemOpen
  \bibfield  {author} {\bibinfo {author} {\bibfnamefont {P}~\bibnamefont
  {Hansch}}\ and\ \bibinfo {author} {\bibfnamefont {L~D}\ \bibnamefont
  {Van~Woerkom}},\ }\bibfield  {title} {\enquote {\bibinfo {title}
  {High-precision intensity-selective observation of multiphoton ionization: a
  new method of photoelectron spectroscopy},}\ }\href@noop {} {\bibfield
  {journal} {\bibinfo  {journal} {Opt. Lett.}\ }\textbf {\bibinfo {volume}
  {21}},\ \bibinfo {pages} {1286--1288} (\bibinfo {year} {1996})}\BibitemShut
  {NoStop}%
\bibitem [{\citenamefont {{International Organization for
  Standardization}}(2005)}]{ISO-11146-profiling-standard}%
  \BibitemOpen
  \bibfield  {author} {\bibinfo {author} {\bibnamefont {{International
  Organization for Standardization}}},\ }\href@noop {} {\emph {\bibinfo {title}
  {ISO 11146-1:2005(E). Lasers and laser-related equipment {-} Test methods for
  laser beam widths, divergence angles and beam propagation ratios {-} Part
  1}}},\ \bibinfo {address} {Geneva, Switzerland} (\bibinfo {year}
  {2005})\BibitemShut {NoStop}%
\bibitem [{\citenamefont {Yariv}(1987)}]{Yariv-quantum-electronics-BOOK}%
  \BibitemOpen
  \bibfield  {author} {\bibinfo {author} {\bibfnamefont {A.}~\bibnamefont
  {Yariv}},\ }\href@noop {} {\emph {\bibinfo {title} {Quantum Electronics}}},\
  \bibinfo {edition} {3rd}\ ed.\ (\bibinfo  {publisher} {Wiley},\ \bibinfo
  {year} {1987})\BibitemShut {NoStop}%
\bibitem [{\citenamefont {Siegman}(1998)}]{Siegman-beam-quality-tutorial}%
  \BibitemOpen
  \bibfield  {author} {\bibinfo {author} {\bibfnamefont {A.~E.}\ \bibnamefont
  {Siegman}},\ }\bibfield  {title} {\enquote {\bibinfo {title} {How to (maybe)
  measure beam quality},}\ }in\ \href@noop {} {\emph {\bibinfo {booktitle}
  {{DPSS} Lasers: Applications and Issues}}},\ \bibinfo {editor} {edited by\
  \bibinfo {editor} {\bibfnamefont {M.~W.}\ \bibnamefont {Dowley}}}\ (\bibinfo
  {publisher} {Optical Society of America},\ \bibinfo {address} {Washington,
  D.C., USA},\ \bibinfo {year} {1998})\ pp.\ \bibinfo {pages}
  {184--199}\BibitemShut {NoStop}%
\end{thebibliography}%

\end{document}